# Climate Sensitivity, Sea Level, and Atmospheric CO$_2$

James Hansen, Makiko Sato, Gary Russell and Pushker Kharecha

NASA Goddard Institute for Space Studies and Columbia University Earth Institute, New York

Cenozoic temperature, sea level and CO$_2$ co-variations provide insights into climate sensitivity to external forcings and sea level sensitivity to climate change. Climate sensitivity depends on the initial climate state, but potentially can be accurately inferred from precise paleoclimate data. Pleistocene climate oscillations yield a fast-feedback climate sensitivity 3 ± 1°C for 4 W/m$^2$ CO$_2$ forcing if Holocene warming relative to the Last Glacial Maximum (LGM) is used as calibration, but the error (uncertainty) is substantial and partly subjective because of poorly defined LGM global temperature and possible human influences in the Holocene. Glacial-to-interglacial climate change leading to the prior (Eemian) interglacial is less ambiguous and implies a sensitivity in the upper part of the above range, i.e., 3-4°C for 4 W/m$^2$ CO$_2$ forcing. Slow feedbacks, especially change of ice sheet size and atmospheric CO$_2$, amplify total Earth system sensitivity by an amount that depends on the time scale considered. Ice sheet response time is poorly defined, but we show that the slow response and hysteresis in prevailing ice sheet models are exaggerated. We use a global model, simplified to essential processes, to investigate state-dependence of climate sensitivity, finding an increased sensitivity towards warmer climates, as low cloud cover is diminished and increased water vapor elevates the tropopause. Burning all fossil fuels, we conclude, would make much of the planet uninhabitable by humans, thus calling into question strategies that emphasize adaptation to climate change.

## 1. Introduction

Humanity is now the dominant force driving changes of Earth's atmospheric composition and climate (IPCC, 2007a). The largest climate forcing today, i.e., the greatest imposed perturbation of the planet's energy balance (IPCC, 2007a; Hansen et al., 2011), is the human-made increase of atmospheric greenhouse gases, especially CO$_2$ from burning of fossil fuels.

Earth's response to climate forcings is slowed by the inertia of the global ocean and the great ice sheets on Greenland and Antarctica, which require centuries, millennia or longer to approach their full response to a climate forcing. This long response time makes the task of avoiding dangerous human alteration of climate particularly difficult, because the human-made climate forcing is being imposed rapidly, with most of the current forcing having been added in just the past several decades. Thus observed climate changes are only a partial response to the current climate forcing, with further response still "in-the-pipeline" (Hansen et al., 1984).

Climate models, numerical climate simulations, provide one way to estimate climate response to forcings, but it is difficult to include realistically all real-world processes. Earth's paleoclimate history allows empirical assessment of climate sensitivity, but the data have large uncertainties. These approaches are usually not fully independent, and the most realistic eventual assessments will be ones combining their greatest strengths.

We use the rich climate history of the Cenozoic era in the oxygen isotope record of ocean sediments to explore the relation of climate change with sea level and atmospheric CO$_2$, inferring climate sensitivity empirically. We use Zachos et al. (2008) isotope data, which are improved over data used in our earlier study (Hansen et al., 2008), and we improve our prescription for separating effects of deep ocean temperature and ice volume in the oxygen isotope record as well as our prescription for relating deep ocean temperature to surface air temperature. Finally, we use an efficient climate model to expand our estimated climate sensitivities beyond the Cenozoic climate range to snowball Earth and runaway greenhouse conditions.

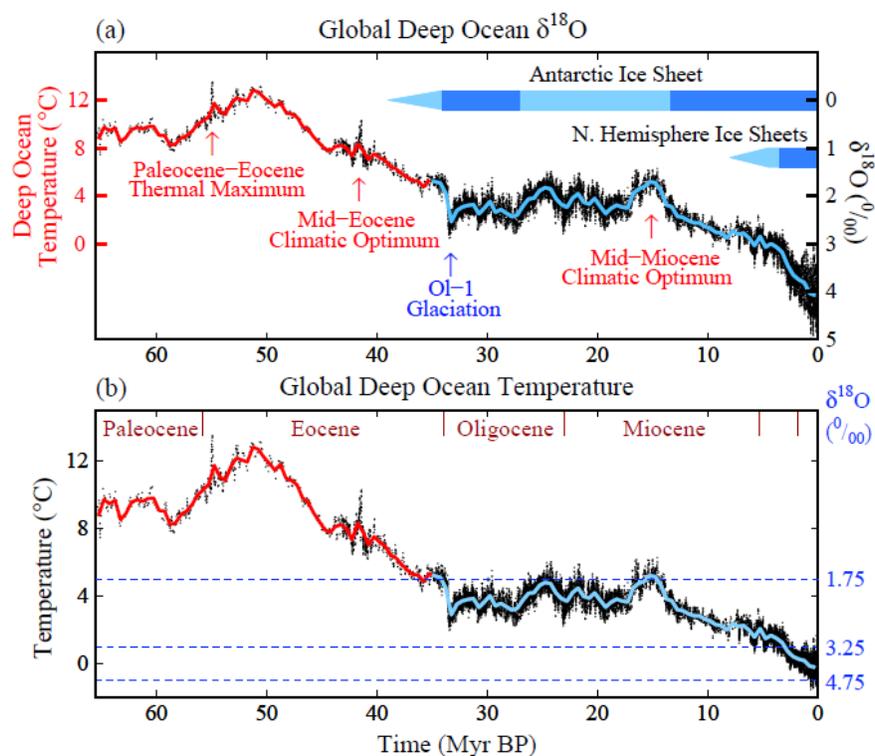

**Fig. 1.** (a) Global deep ocean $\delta^{18}O$ from Zachos et al. (2008) and (b) estimated deep ocean temperature based on the prescription in our present paper. Black data points are 5-point running means of original temporal resolution; red and blue curves have 500 kyr resolution. Coarse temporal sampling reduces the amplitude of glacial-interglacial oscillations in intervals 7-17, 35-42 and 44-65 Myr BP.

## 2. Overview of Cenozoic Climate and Our Analysis Approach

The Cenozoic era, the past 65.5 Myr (million years), provides a valuable perspective on climate (Zachos et al., 2001; Hansen et al., 2008) and sea level change (Gasson et al., 2012), and Cenozoic data help clarify our analysis approach. The principal data set we use is the temporal variation of the oxygen isotope ratio ($\delta^{18}O$ relative to $\delta^{16}O$, Fig. 1a right scale) in the shells of deep-ocean-dwelling microscopic shelled animals (foraminifera) in a near-global compilation of ocean sediment cores (Zachos et al., 2008). $\delta^{18}O$ yields an estimate of deep ocean temperature (Fig. 1b), as discussed in section 3. Note that coarse temporal resolution of $\delta^{18}O$ data in intervals 7-17, 35-42 and 44-65 Myr reduces the apparent amplitude of glacial-interglacial climate fluctuations (see Fig. S1, Supplementary Material). We use additional proxy measures of climate change to supplement the $\delta^{18}O$ data in our quantitative analyses.

Carbon dioxide is involved in climate change throughout the Cenozoic era, both as a climate forcing and as a climate feedback. Long-term Cenozoic temperature trends, the warming up to about 50 Myr BP (before present) and subsequent long-term cooling, are likely to be, at least in large part, a result of the changing natural source of atmospheric $CO_2$, which is volcanic emissions that occur mainly at continental margins due to plate tectonics (popularly "continental drift"); tectonic activity also affects the weathering sink for $CO_2$ by exposing fresh rock. The $CO_2$ tectonic source grew from 60 to 50 My BP as India subducted carbonate-rich ocean crust while moving through the present Indian Ocean prior to its collision with Asia about 50 Myr BP (Kent and Muttoni, 2008), causing atmospheric $CO_2$ to reach levels of the order of 1000 ppm (parts per million) at 50 Myr BP (Beerling & Royer, 2011). Since then, atmospheric $CO_2$ declined as the Indian and Atlantic Oceans have been major depocenters for carbonate and organic sediments while subduction of carbonate-rich crust has been limited mainly to small regions near Indonesia and Central America (Edmond and Huh, 2003), thus allowing $CO_2$ to decline to levels as low as 170 ppm during recent glacial periods (Petit et al., 1999). Climate forcing due to $CO_2$ change from 1000 ppm to 170 ppm is more than 10 W/m$^2$, which compares with forcings of the order of 1 W/m$^2$ for competing climate forcings during the Cenozoic (Hansen et al., 2008), specifically long-term change of solar irradiance and change of planetary albedo (reflectance) due to the overall minor displacement of continents in that era.

Superimposed on the long-term trends are occasional global warming spikes, "hyperthermals", most prominently the Paleocene-Eocene Thermal Maximum (PETM) at ~ 56 Myr BP (Kennett & Stott, 1991) and the Mid-Eocene Climatic Optimum (MECO) at ~ 42 Myr BP (Bohaty et al., 2009), coincident with large temporary increases of atmospheric $CO_2$. The most studied hyperthermal, the PETM, caused global warming of at least 5°C coincident with injection of a likely 4000-7000 Gt of isotopically light carbon into the atmosphere and ocean (Dunkley Jones et al., 2010). The size of the carbon injection is estimated from changes in the stable carbon isotope ratio $^{13}C/^{12}C$ in sediments and from ocean acidification implied by changes in the ocean depth below which carbonate dissolution occurred.

The potential carbon source for hyperthermal warming that received most initial attention was methane hydrates on continental shelves, which could be destabilized by sea floor warming (Dickens et al., 1995). Alternative sources include release of carbon from Antarctic permafrost and peat (DeConto et al., 2012). Regardless of the carbon source(s), it has been shown that the hyperthermals were astronomically paced, spurred by coincident maxima in Earth's orbit eccentricity and spin axis tilt (Lourens et al., 2005), which increased high latitude insolation and warming. The PETM was followed by successively weaker astronomically-paced hyperthermals, suggesting that the carbon source(s) partially recharged in the interim (Lunt et al., 2011). A high temporal resolution sediment core from the New Jersey continental shelf (Sluijs et al., 2007) reveals that PETM warming in at least that region began about 3000 years prior to a massive release of isotopically light carbon. This lag and climate simulations (Lunt et al., 2010a) that produce large warming at intermediate ocean depths in response to initial surface warming are consistent with the concept of a methane hydrate role in hyperthermal events.

The hyperthermals confirm understanding about the long recovery time of Earth's carbon cycle (Archer, 2005) and reveal the potential for threshold or "tipping point" behavior with large amplifying climate feedback in response to warming (Thomas et al., 2002). One implication is that if humans burn most of the fossil fuels, thus injecting into the atmosphere an amount of $CO_2$ at least comparable to that injected during the PETM, the $CO_2$ would stay in the surface carbon reservoirs (atmosphere, ocean, soil, biosphere) for tens of thousands of years, long enough for the atmosphere, ocean, and ice sheets to fully respond to the changed atmospheric composition. In addition, there is the potential that global warming from fossil fuel $CO_2$ could spur release of $CH_4$ and $CO_2$ from methane hydrates or permafrost. Carbon release during the hyperthermals required several thousand years, but that long injection time may have been a function of the pace of the astronomical forcing, which is much slower than the pace of fossil fuel burning.

The Cenozoic record also reveals the amplification of climate change that occurs with growth or decay of ice sheets, as is apparent at about 34 Myr BP when Earth became cool enough for large scale glaciation of Antarctica and in the most recent 3-5 Myr with growth of Northern Hemisphere ice sheets. Global climate fluctuated in the 20 Myr following Antarctic glaciation with warmth during the Mid-Miocene Climatic Optimum (MMCO, 15 Myr BP) possibly comparable to that at 34 Myr BP, as, e.g., Germany became warm enough to harbor snakes and crocodiles that require annual temperature about 20°C or higher and winter temperature above 10°C (Ivanov and Bohme, 2011). Antarctic vegetation in the MMCO implies summer temperature about 11°C warmer than today (Feakins et al., 2012) and annual sea surface temperatures ranging from 0 to 11.5°C (Warny et al., 2009).

Superimposed on the long-term trends, in addition to occasional hyperthermals, are continual high frequency temperature oscillations, which are apparent in Fig. 1 after 34 Myr BP, when Earth became cold enough for a large ice sheet to form on Antarctica, and are still more prominent during ice sheet growth in the Northern Hemisphere. These climate oscillations have dominant periodicities, ranging from about 20 to 400 kyr, that coincide with variations of Earth's orbital elements (Hays et al., 1976), specifically the tilt of Earth's spin axis, the eccentricity of the orbit, and the time of year when Earth is closest to the Sun. The slowly changing orbit and tilt of the spin axis affect the seasonal distribution of insolation (Berger, 1978), and thus the growth and decay of ice sheets, as proposed by Milankovitch (1941). Atmospheric $CO_2$, $CH_4$, and $N_2O$ vary almost synchronously with global temperature during the last 800,000 years when precise data are available from ice cores, the GHGs providing an amplifying feedback that magnifies the climate change instigated by orbit perturbations (Jouzel et al., 2007; Kohler et al., 2010; Masson-Delmotte et al., 2010).

Ocean and atmosphere dynamical effects have been suggested as possible causes of some climate change within the Cenozoic, e.g., topographical effects of mountain building (Ruddiman et al., 1989), closing of the Panama Seaway (Keigwin, 1982), or opening of the Drake Passage (Kennett, 1977). Climate modeling studies with orographic changes confirm significant effects on monsoons and on Eurasian temperature (Ramstein et al., 1997). Modeling studies indicate that closing of the Panama Seaway results in a more intense Atlantic thermohaline circulation, but only small effects on Northern Hemisphere ice sheets (Lunt et al., 2008). Opening of the Drake Passage surely affected ocean circulation around Antarctica, but efforts to find a significant effect on global temperature have relied on speculation about possible effects on atmospheric $CO_2$ (Scher & Martin, 2006). Overall there is no strong evidence that dynamical effects are a major direct contributor to Cenozoic global temperature change.

We hypothesize that the global climate variations of the Cenozoic (Fig. 1) can be understood and analyzed via slow temporal changes of Earth's energy balance, which is a function of solar irradiance, atmospheric composition (specifically long-lived GHGs), and planetary surface albedo. Using measured amounts of GHGs during the past 800,000 years of glacial-interglacial climate oscillations and surface albedo inferred from sea level data, we show that a single empirical "fast-feedback" climate sensitivity can account well for global temperature change over that range of climate states. It is certain that over a large climate range climate sensitivity must become a strong function of climate state, and thus we use a simplified climate model to investigate the dependence of climate sensitivity on climate state. Finally we use our estimated state-dependent climate sensitivity to infer Cenozoic $CO_2$ change and compare this with proxy $CO_2$ data, focusing on the Eocene climatic optimum, the Oligocene glaciation, Miocene optimum, and the Pliocene.

### 3. Deep Ocean Temperature and Sea Level in the Cenozoic Era

The $\delta^{18}O$ stable isotope ratio was the first paleothermometer, proposed by Urey (1947) and developed especially by Emiliani (1955). There are now several alternative proxy measures of ancient climate change, but the $\delta^{18}O$ data (Fig. 1a) of Zachos et al. (2008), a conglomerate of global ocean sediment cores, is well-suited for our purpose as it covers the Cenozoic era with good temporal resolution. There are large, even dominant, non-climatic causes of $\delta^{18}O$ changes over hundreds of millions of years (Jaffres et al., 2007), but non-climatic change may be small in the past few hundred million years (Wallman, 2001) and is generally neglected in Cenozoic climate studies. The principal difficulty in using the $\delta^{18}O$ record to estimate global deep ocean temperature, in the absence of non-climatic change, is that $\delta^{18}O$ is affected by global ice mass as well as deep ocean temperature.

We make a simple estimate of global sea level change for the Cenozoic era using the near-global $\delta^{18}O$ compilation of Zachos et al. (2008). More elaborate and accurate approaches, including use of models, will surely be devised, but comparison of our result with other approaches is instructive regarding basic issues such as the vulnerability of today's ice sheets to near-term global warming and the magnitude of hysteresis effects in ice sheet growth and decay.

During the early Cenozoic, between 65.5 and 35 Myr BP, Earth was so warm that there was little ice on the planet and deep ocean temperature is approximated by (Zachos et al., 2001),

$$T_{do}(°C) = -4 \delta^{18}O + 12 \qquad (\text{for } \delta^{18}O < 1.75). \qquad (1)$$

Hansen et al. (2008) made the approximation that, as Earth became colder and continental ice sheets grew, further increase of $\delta^{18}O$ was due in equal parts to deep ocean temperature change and ice mass change,

$$T_{do}(°C) = -2 (\delta^{18}O - 4.25) \qquad (\text{for } \delta^{18}O > 1.75). \qquad (2)$$

Equal division of the $\delta^{18}O$ change into temperature change and ice volume change was suggested by comparing $\delta^{18}O$ at the endpoints of the climate change from the nearly ice-free planet at 35 Myr BP (when $\delta^{18}O \sim 1.75$) to the Last Glacial Maximum (LGM), which peaked ~ 20 kyr BP. The change of $\delta^{18}O$ between these two extreme climate states (~ 3), is twice the change of $\delta^{18}O$ due to temperature change alone (~ 1.5), with temperature change based on the linear relation (1) and estimates of $T_{do} \sim 5°C$ at 35 Myr BP (Fig. 1) and ~ -1°C at the LGM (Cutler et al., 2003).

This approximation can easily be made more realistic. Although ice volume and deep ocean temperature changes contributed comparable amounts to $\delta^{18}O$ change on average over the full range from 35 Myr to 20 kyr BP, the temperature change portion of $\delta^{18}O$ change must decrease as the deep ocean temperature approaches the freezing point (Waelbroeck et al., 2002). The rapid increase of $\delta^{18}O$ in the past few million years was associated with appearance of Northern Hemisphere ice sheets, symbolized by the dark blue bar in Fig. 1a.

Sea level change between the LGM and Holocene was ~120 m (Fairbanks, 1989; Peltier & Fairbanks, 2006). Thus two-thirds of the 180 m sea level change between the ice-free planet and LGM occurred with formation of Northern Hemisphere ice (and probably some increased volume of Antarctic ice). Thus rather than taking the 180 m sea level change between the nearly ice-free planet of 34 Myr BP and the LGM as being linear over the entire range (with 90 m for $\delta^{18}O < 3.25$ and 90 m for $\delta^{18}O > 3.25$), it is more realistic to assign 60 m of sea level change to $\delta^{18}O$ 1.75-3.25 and 120 m to $\delta^{18}O > 3.25$. The total deep ocean temperature change of 6°C for change of $\delta^{18}O$ from 1.75 to 4.75 is then divided two-thirds (4°C) for the $\delta^{18}O$ range 1.75-3.25 and 2°C for the $\delta^{18}O$ range 3.25-4.75. Algebraically

$$SL (m) = 60 - 40 (\delta^{18}O - 1.75) \qquad (\text{for } \delta^{18}O < 3.25) \qquad (3)$$

$$SL (m) = -120 (\delta^{18}O - 3.25)/1.65 \qquad (\text{for } \delta^{18}O > 3.25) \qquad (4)$$

$$T_{do}(°C) = 5 - 8(\delta^{18}O - 1.75)/3 \qquad (\text{for } \delta^{18}O < 3.25) \qquad (5)$$

$$T_{do}(°C) = 1 - 4.4 (\delta^{18}O - 3.25)/3 \qquad (\text{for } \delta^{18}O > 3.25) \qquad (6)$$

where SL is sea level and its zero point is the late Holocene level. The coefficients in equation (4) and (5) account for the fact that the mean LGM value of $\delta^{18}O$ is ~ 4.9. The resulting deep ocean temperature is shown in Fig. 1(b) for the full Cenozoic era.

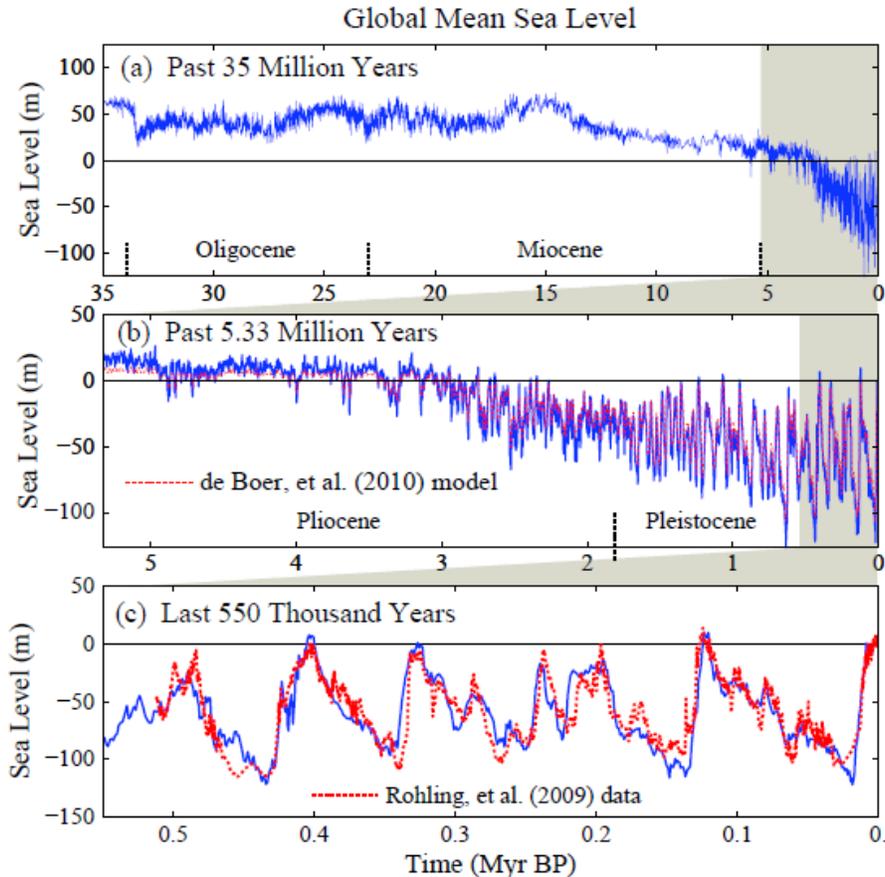

**Fig. 2.** Sea level from equations (3) and (4) using $\delta^{18}O$ data of Zachos et al. (2008), compared in (b) with ice sheet model results of de Boer et al. (2010) and in (c) to the sea level analysis of Rohling et al. (2009).

Sea level from equations (3) and (4) is shown by the blue curves in Fig. 2, including comparison (Fig. 2c) with the late Pleistocene sea level record of Rohling et al. (2009), which is based on analysis of Red Sea sediments, and comparison (Fig. 2b) with the sea level chronology of de Boer et al. (2010), which is based on ice sheet modeling with the $\delta^{18}O$ data of Zachos et al. (2008) as a principal input driving the ice sheet model. Comparison of our result with that of de Boer et al. (2010) for the other periods of Fig. 2 is included in our Supplementary Material, where we also make available our numerical data. Deep ocean temperature from equations 5 and 6 is shown for the Pliocene and Pleistocene in Fig. 3 and for the entire Cenozoic era in Fig. 1.

Differences between our inferred sea level chronology and that from the ice sheet model (de Boer et al., 2010) are relevant to assessment of the potential danger to humanity of future sea level rise. Our estimated sea levels reach +5-10 m above present sea level during recent interglacial periods that were barely warmer than the Holocene, while the ice sheet model yields maxima at most ~ 1m above current sea level. We find Pliocene sea level varying between about +20 m and –50 m, with early Pliocene averaging about +15 m; the ice sheet model has less variable sea level with the early Pliocene averaging about +8 m. A 15 m sea level rise implies that the East Antarctic ice sheet, as well as West Antarctica and Greenland ice, were unstable at a global temperature no higher than those projected to occur this century (IPCC, 2007).

How can we interpret these differences, and what is the merit of our simple $\delta^{18}O$ scaling? Ice sheet models constrained by multiple observations may eventually provide our best estimate of sea level change, but as yet models are primitive. Hansen (2005, 2007) argues that real ice sheets are more responsive to climate change than is found in most ice sheet models. Our simple scaling approximation implicitly assumes that ice sheets are sufficiently responsive to climate change that hysteresis is not a dominant effect; in other words ice volume on millennial time scales is a function of temperature and does not depend much on whether Earth is in a warming or cooling phase. Thus our simple transparent calculation may provide a useful comparison with geologic data for sea level change and with results of ice sheet models.

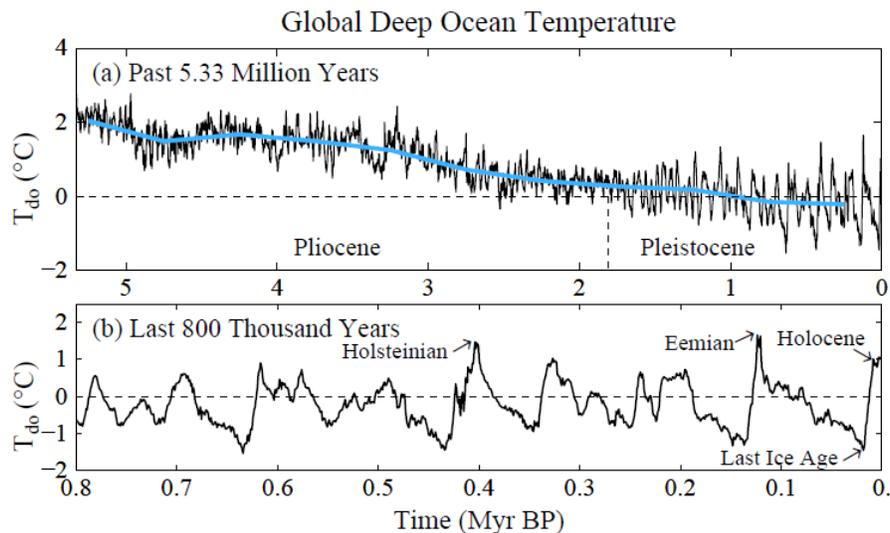

**Fig. 3.** Deep ocean temperature in (a) the Pliocene and Pleistocene, and (b) the last 800,000 years. High frequency variations (black) are 5-point running means of original data (Zachos et al., 2008), while the blue curve has 500 kyr resolution. Deep ocean temperature for the entire Cenozoic era is in Fig. 1 (b).

    We cannot a priori define accurately the error in our sea level estimates, but we can compare with geologic data in specific cases as a check on reasonableness. Our results (Fig. 2) yield two instances in the past million years when sea level reached heights well above current sea level: +9.8 m in the Eemian (~120 kyr BP, also known as Marine Isotope Stage 5e or MIS 5e) and +7.1 m in the Holsteinian (~400 kyr BP, also known as MIS 11). Indeed, these are the two interglacial periods in the late Pleistocene that traditional geologic methods identify as likely having sea level exceeding that in the Holocene. Geologic evidence, mainly coral reefs on tectonically stable coasts, was described in the review of Overpeck et al. (2006) as favoring an Eemian maximum of +4 to >6 m. Rohling et al. (2008) cite many studies concluding that mean sea level was 4-6 m above current sea level during the warmest portion of the Eemian, 123-119 ky BP, note that several of these studies suggest Eemian sea level fluctuations up to +10 m, and provide the first continuous sea level data supporting rapid Eemian sea level fluctuations. Kopp et al. (2009) made a statistical analysis of data from a large number of sites, concluding that there was 95% probability that Eemian sea level reached at least +6.6 m and 67% probability that it exceeded 8 m.

    Holsteinian sea level is more difficult to reconstruct from geologic data because of its age, and there has been a long-standing controversy concerning a substantial body of geological shoreline evidence for a +20 m late Holsteinian sea level that Hearty and colleagues have found on numerous sites (Hearty et al., 1999; Hearty, 2010; numerous references pro and con are contained in the references provided in our present paragraph). Rohling et al. (2010) note that their temporally continuous Red Sea record "strongly supports the MIS-11 sea level review of Bowen (2010), which also places MIS-11 sea level within uncertainties at the present-day level." This issue is important because both ice core data (Jouzel et al., 2007) and ocean sediment core data (see below) indicate that the Holsteinian period was only moderately warmer than the Holocene with similar Earth orbital parameters. We suggest that the resolution of this issue is consistent with our estimate of ~ +7 m Holsteinian global sea level, and is provided by Raymo

and Mitrovica (2012), who pointed out the need to make a glacial isostatic adjustment (GIA) correction for post-glacial crustal subsidence at the places where Hearty and others deduced local sea level change. The uncertainties in GIA modeling lead Raymo and Mitrovica (2012) to conclude that peak Holsteinian global sea level was in the range +6-13 m relative to the present. Thus, it seems to us, there is a reasonable resolution of the long-standing Holsteinian controversy, with substantial implications for humanity, as discussed in later sections.

We now address differences between our sea level estimates and those from ice sheet models. We refer to both the 1-D ice sheet modeling of de Boer et al. (2010), which was used to calculate sea level for the entire Cenozoic era, and the 3-D ice sheet model of Bintanja et al. (2005), which was used for simulations of the past million years. The differences most relevant to humanity occur in the interglacial periods slightly warmer than the Holocene, including the Eemian and Hosteinian, as well as the Pliocene, which may have been as warm as projected for later this century. Both the 3-D model of Bintanja et al. (2005) and the 1-D model of de Boer et al. (2010) yield maximum Eemian and Hosteinian sea levels of ~ 1 m relative to the Holocene. de Boer et al (2010) obtain ~ +8 m for the early Pliocene, which compares with our ~ +15 m.

These differences reveal that the modeled ice sheets are less susceptible to change in response to global temperature variation compared to our $\delta^{18}O$ analysis. Yet the ice sheet models do a good job of reproducing sea level change for climates colder than the Holocene, as shown in Fig. 2 and Fig. S2. One possibility is that the ice sheet models are too lethargic for climates warmer than the Holocene. Hansen and Sato (2012) point out the sudden change in the responsiveness of the ice sheet model of Bintanja et al. (2005) when sea level reaches today's level (see Figs. 3 and 4 of Hansen and Sato, 2012) and they note that the empirical sea level data provide no evidence of such a sudden change. The explanation conceivably lies in the fact that the models have many parameters and their operation includes use of "targets" (de Boer et al., 2010) that affect the model results, because these choices might yield different results for warmer climates than the results for colder climates. Because of the potential that model development choices might be influenced by expectations of a "correct" result, it is useful to have estimates independent of the models based on alternative assumptions.

Note that our approach also involves "targets" based on expected behavior, albeit simple transparent ones. Our two-legged linear approximation of sea level (Equations 3 and 4) assumes that sea level in the LGM was 120 m lower than today and that sea level was 60 m higher than today 35 Myr BP. This latter assumption may need to be adjusted if glaciers and ice caps in the Eocene had a volume of tens of meters of sea level. However, Miller et al. (2012) conclude that there was a sea level fall of ~ 55 m at the Eocene-Oligocene transition, consistent with our assumption that Eocene ice probably did not contain more than ~ 10 m of sea level.

Real world data for Earth's sea level history ultimately must provide assessment of sea level sensitivity to climate change. A recent comprehensive review (Gasson et al., 2012) reveals there are still wide uncertainties about Earth's sea level history that are especially large for time scales of tens of millions of years or longer, which is long enough for substantial changes of the shape and volume of ocean basins. Gasson et al. (2012) plot regional (New Jersey) sea level (their Fig. 14) against deep ocean temperature inferred from the magnesium/calcium ratio (Mg/Ca) of deep ocean foraminifera (Lear et al., 2000), finding evidence for a non-linear sea level response to temperature roughly consistent with the modeling of de Boer et al. (2010). Sea level change is limited for Mg/Ca temperatures up to about 5°C above current values, whereupon a rather abrupt sea level rise of several tens of meters occurs, presumably representing the loss of Antarctic ice. However, the uncertainty in the reconstructed sea level is tens of meters and the uncertainty in the Mg/Ca temperature is sufficient to encompass the result from our $\delta^{18}O$ prescription, which has comparable contributions of ice volume change and deep ocean temperature change at the late Eocene glaciation of Antarctica.

Furthermore, the potential sea level rise of most practical importance is the first 15 m above the Holocene level. It is such "moderate" sea level change for which we particularly question the projections implied by current ice sheet models. Empirical assessment depends upon real world sea level data in periods warmer than the Holocene. There is strong evidence, discussed above, that sea level was several meters higher in recent warm interglacial periods, consistent with our data interpretation. The Pliocene provides data extension to still warmer climates. Our interpretation of $\delta^{18}O$ data suggests that early Pliocene sea level change (due to ice volume change) reached about +15 m, and it also indicates sea level fluctuations as large as 20-40 m. Sea level data for mid-Pliocene warm periods, of comparable warmth to average early Pliocene conditions (Fig. 3), suggest sea heights as great as +15-25 m (Dowsett et al., 1999; Dwyer and Chandler, 2009). Miller et al. (2012) find a Pliocene sea level maximum 22 ± 10 m (95% confidence). Glacial isostatic adjustment creates uncertainty in sea level reconstructions based on shoreline geologic data (Raymo et al., 2011), which could be reduced via appropriately distributed field studies. Dwyer and Chandler (2009) separate Pliocene ice volume and temperature in deep ocean $\delta^{18}O$ via ostracode Mg/Ca temperatures, finding sea level maxima and oscillations comparable to our results. Altogether the empirical data provide strong evidence against the lethargy and strong hysteresis effects of at least some ice sheet models.

## 4. Surface Air Temperature Change

The temperature of most interest to humanity is surface air temperature. A record of past global surface temperature is needed for empirical inference of global climate sensitivity. Given that climate sensitivity can depend on the initial climate state and on the magnitude and sign of the climate forcing, a continuous record of global temperature over a wide range of climate states would be especially useful. Because of the singularly rich climate story in Cenozoic deep ocean $\delta^{18}O$ (Fig. 1), unrivaled in detail and self-consistency by alternative climate proxies, we use deep ocean $\delta^{18}O$ to provide the fine structure of Cenozoic temperature change. We use surface temperature proxies from the LGM, the Pliocene and the Eocene to calibrate and check the relation between deep ocean and surface temperature change.

The temperature signal in deep ocean $\delta^{18}O$ refers to the sea surface where cold dense water formed and sank to the ocean bottom, the principal location of deep water formation being the Southern Ocean. Empirical data and climate models concur that surface temperature change is generally amplified at high latitudes, which tends to make temperature change at the site of deep water formation an overestimate of global temperature change. Empirical data and climate models also concur that surface temperature change is amplified over land areas, which tends to make temperature change at the site of deep water an underestimate of global temperature. Hansen et al. (2008) and Hansen and Sato (2012) noted that these two factors were substantially offsetting, and thus they made the assumption that benthic foraminifera provide a good approximation of global mean temperature change for most of the Cenozoic era.

However, this approximation breaks down in the late Cenozoic for two reasons. First, the deep ocean and high latitude surface ocean where deep water forms are approaching the freezing point in the late Cenozoic. As Earth's surface cools further, cold conditions spread to lower latitudes but polar surface water and the deep ocean cannot become much colder, and thus the benthic foraminifera record a temperature change smaller than the global average surface temperature change (Waelbroeck et al., 2002). Second, the last 5.33 Myr of the Cenozoic, the Pliocene and Pleistocene, were the time that global cooling reached a degree such that large ice sheets could form in the Northern Hemisphere. When a climate forcing, or a slow climate feedback such as ice sheet formation, occurs in one hemisphere, the temperature change is much larger in the hemisphere with the forcing (cf. examples in Hansen et al., 2005). Thus cooling during the last 5.33 Myr in the Southern Ocean site of deep water formation was smaller than the global average cooling.

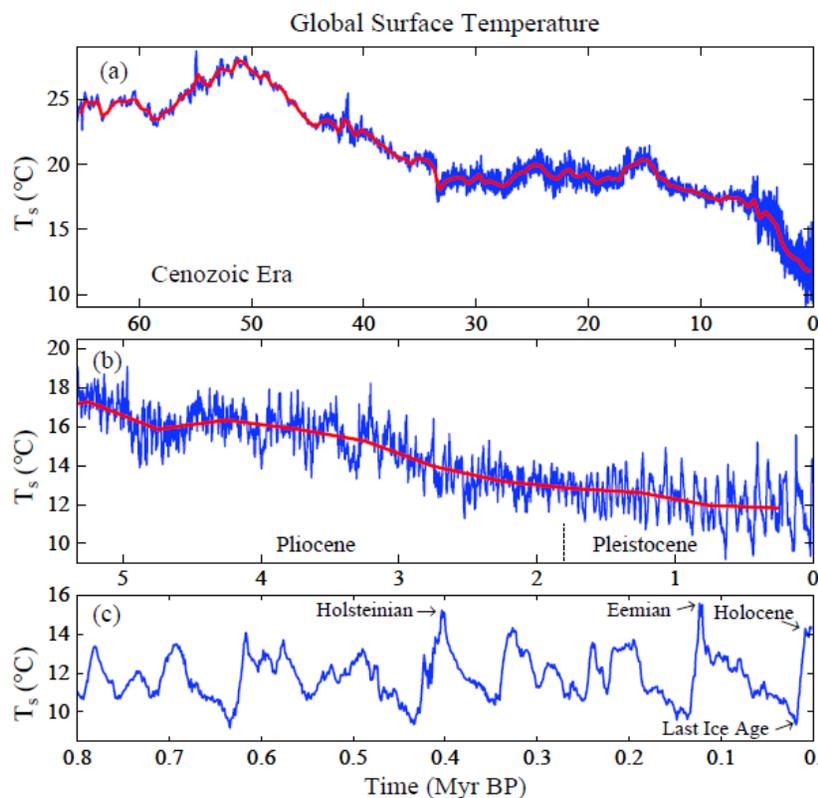

**Fig. 4.** Surface temperature estimate for the past 65.5 million years, including an expanded time scale for (b) the Pliocene and Pleistocene and (c) the last 800,000 years. The red curve has 500 kyr resolution. Data for this and other figures are available in our Supplementary Material.

We especially want our global surface temperature reconstruction to be accurate for the Pliocene and Pleistocene because the global temperature changes that are expected by the end of this century, if humanity continues to rapidly change atmospheric composition, are of a magnitude comparable to climate change in those epochs (IPCC, 2007). Fortunately, sufficient information is available on surface temperature change in the Pliocene and Pleistocene to allow us to scale the deep ocean temperature change by appropriate factors, thus retaining the temporal variations in the $\delta^{18}O$ while also having a realistic magnitude for the total temperature change over these epochs.

Pliocene temperature is known quite well because of a long-term effort to reconstruct the climate conditions during the mid-Pliocene warm period (3.29-2.97 Myr BP) and a coordinated effort to numerically simulate the climate by many modeling group (Haywood et al., 2010 and papers referenced therein). The reconstructed Pliocene climate used data for the warmest conditions found in the mid-Pliocene period, which would be similar to average conditions in the early Pliocene (Fig. 3). These boundary conditions were used by eight modeling groups to simulate Pliocene climate with atmospheric general circulation models. Although atmosphere-ocean models have difficulty replicating Pliocene climate, atmospheric models forced by specified surface boundary conditions are expected to be capable of calculating global surface temperature with reasonable accuracy. The eight global models yield Pliocene global warming of $3 \pm 1°C$ relative to the Holocene (Haywood et al., 2012). This Pliocene warming is an amplification by a factor of 2.5 of the deep ocean temperature change.

Similarly, for the reasons given above, the deep ocean temperature change of 2.25°C between the Holocene and the Last Glacial Maximum (LGM) is surely an underestimate of the surface air temperature change. Unfortunately, there is a wide range of estimates for LGM cooling, approximately 3-6°C, as discussed in Section 6. Thus we take 4.5°C as our best estimate for LGM cooling, implying an amplification of surface temperature change by a factor of two relative to deep ocean temperature change for this climate interval.

We obtain an absolute temperature scale using the Jones et al. (1999) estimate of 14°C as the global mean surface temperature for 1961-1990, which corresponds to ~13.9°C for the 1951-1980 base period that we normally employ (Hansen et al., 2010) and ~14.4°C for the first decade of the 21st century. We attach the instrumental temperature record to the paleo data by assuming that the first decade of the 21st century exceeds the Holocene mean by 0.25 ± 0.25°C. Global temperature probably declined over the past several millennia (Mayewski et al., 2004), but we suggest that warming of the past century has brought global temperature to a level that now slightly exceeds the Holocene mean, judging from sea level trends and ice sheet mass loss. Sea level is now rising 3.1 mm/yr or 3.1m/millennium (Nerem et al., 2006), an order of magnitude faster than the rate during the past several thousand years, and Greenland and Antarctica are losing mass at accelerating rates (Rignot et al., 2011; King et al., 2012). Our assumption that global temperature passed the Holocene mean a few decades ago is consistent with the rapid change of ice sheet mass balance in the past few decades (Zwally et al., 2011). The above concatenation of instrumental and paleo records yields a Holocene mean of 14.15°C and Holocene maximum (from 5-point smoothed $\delta^{18}O$) of 14.3°C at 8.6 kyr BP.

Given Holocene temperature 14.15°C and LGM cooling of 4.5°C, Early Pliocene mean temperature 3°C warmer than the Holocene leads to the prescription:

$$T_S(°C) = 2 \times T_{do} + 12.25°C \qquad \text{(Pleistocene)} \qquad (7)$$

$$T_S(°C) = 2.5 \times T_{do} + 12.15°C \qquad \text{(Pliocene)} \qquad (8)$$

This prescription yields a maximum Eemian temperature 15.56°C, which is ~1.4°C warmer than the Holocene mean and ~1.8°C warmer than the 1880-1920 mean. Clark and Huybers (2009) fit a polynomial to proxy temperatures for the Eemian, finding warming as much as +5°C at high northern latitudes but global warming of +1.7°C "relative to the present interglacial before industrialization". Other analyses of Eemian data find global sea surface temperature warmer than late Holocene by 0.7 ± 0.6°C (McKay et al., 2011) and all-surface warming of 2°C (Turney and Jones, 2010), all in reasonable accord with our prescription.

Our first estimate of global temperature for the remainder of the Cenozoic assumes that $\Delta T_s = \Delta T_{do}$ prior to 5.33 Myr BP, i.e., prior to the Plio-Pleistocene, which yields peak $T_s$ ~28°C at 50 Myr BP (Fig. 4). This is at the low end of the range of current multi-proxy measures of sea surface temperature for the Early Eocene Climatic Optimum (Zachos et al., 2006; Pearson et al., 2007; Hollis et al., 2009). Climate models are marginally able to reproduce this level of Eocene warmth, but the models require extraordinarily high $CO_2$ levels, e.g., 2240-4480 ppm (Hollis et al., 2012) and 2500-6500 ppm (Lunt et al., 2012), and the quasi-agreement between data and models requires an assumption that some of the proxy temperatures are biased toward summer values. Moreover, taking the proxy sea surface temperature data for peak Eocene period (55-48 Myr BP) at face value yields a global temperature of 33-34°C (Fig. 3 of Bijl et al., 2009), which would require an even larger $CO_2$ amount with the same climate models. Thus below we also consider the implications for climate sensitivity of an assumption that $\Delta T_s = 1.5 \times \Delta T_{do}$ prior to 5.33 Myr BP, which yields $T_s$ ~33°C at 50 Myr BP (Fig. S3).

## 5. Climate Sensitivity

Climate sensitivity (S) is the equilibrium global surface temperature change ($\Delta T_{eq}$) in response to a specified unit forcing after the planet has come back to energy balance,

$$S = \Delta T_{eq}/F, \qquad (7)$$

i.e., climate sensitivity is the eventual (equilibrium) global temperature change per unit forcing. Climate sensitivity depends upon climate feedbacks, the many physical processes that come into play as climate changes in response to a forcing. Positive (amplifying) feedbacks increase the climate response, while negative (diminishing) feedbacks reduce the response.

We usually discuss climate sensitivity in terms of global mean temperature response to a 4 W/m$^2$ CO$_2$ forcing. One merit of this standard forcing is that its magnitude is similar to anticipated near-term human-made climate forcing, thus avoiding the need to continually scale the unit sensitivity to achieve an applicable magnitude. A second merit is that the efficacy of forcings varies from one forcing mechanism to another (Hansen et al., 2005), so it is useful to employ the forcing mechanism of greatest interest. Finally, the 4 W/m$^2$ CO$_2$ forcing avoids the uncertainty in the exact magnitude of a doubled CO$_2$ forcing [IPCC (2007) estimate 3.7 W/m$^2$ for doubled CO$_2$ while Hansen et al. (2005) obtain 4.1 W/m$^2$], as well as problems associated with the fact that doubled CO$_2$ forcing varies as the CO$_2$ amount changes [the assumption that each CO$_2$ doubling has the same forcing is meant to approximate the effect of CO$_2$ absorption line saturation, but actually the forcing per doubling increases as CO$_2$ increases (Hansen et al., 2005; Colman and McAvaney, 2009)].

Climate feedbacks are the core of the climate problem. Climate feedbacks can be confusing, because in climate analyses what is sometimes a climate forcing is other times a climate feedback. A CO$_2$ decrease from say ~1000 ppm in the early Cenozoic to 170-300 ppm in the Pleistocene, caused by shifting plate tectonics, is a climate forcing, a perturbation of Earth's energy balance that alters the temperature. Glacial-interglacial oscillations of CO$_2$ amount and ice sheet size are both slow climate feedbacks, because glacial-interglacial climate oscillations largely are instigated by insolation changes as Earth's orbit and tilt of its spin axis change, with the climate change then amplified by nearly coincident change of CO$_2$ amount and surface albedo. However, for the sake of analysis we can also choose and compare periods that are in quasi-equilibrium, periods during which there was little change of ice sheet size or GHG amount. For example, we can compare conditions averaged over several millennia in the LGM with mean Holocene conditions. Earth's average energy imbalance within each of these periods had to be a small fraction of 1 W/m$^2$. Such a planetary energy imbalance is very small compared to the boundary condition "forcings", such as changed GHG amount and changed surface albedo that maintain the glacial-to-interglacial climate change.

*(a) Fast-Feedback Sensitivity: LGM-Holocene*

The average fast-feedback climate sensitivity over the LGM-Holocene range of climate states can be assessed by comparing estimated global temperature change and climate forcing change between those two climate states (Hansen et al., 1984; Lorius et al., 1990). The appropriate climate forcings are the changes of long-lived GHGs and surface properties on the planet. Fast feedbacks include water vapor, clouds, aerosols, and sea ice changes.

This fast-feedback sensitivity is relevant to estimating the climate impact of human-made climate forcings, because the size of ice sheets is not expected to change significantly in decades or even in a century and GHGs can be specified as a forcing. GHGs change in response to climate change, but it is common to include these feedbacks as part of the climate forcing by using observed GHG changes for the past and calculated GHGs for the future, with calculated amounts based on carbon cycle and atmospheric chemistry models.

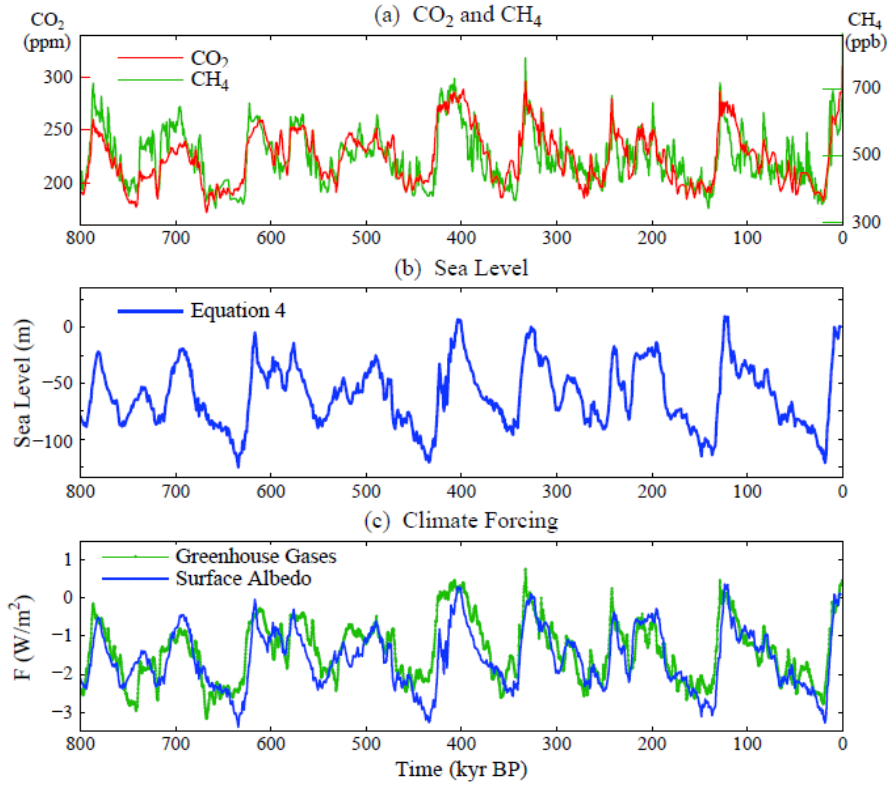

**Fig. 5.** (a) $CO_2$ and $CH_4$ from ice cores, (b) sea level from equation (4), and (c) resulting climate forcings (see text).

Climate forcings due to past changes of GHGs and surface albedo can be computed for the last 800,000 years using data from polar ice cores and ocean sediment cores. We use $CO_2$ (Luthi et al., 2008) and $CH_4$ (Loulergue et al., 2008) data from Antarctic ice cores (Fig. 5a) to calculate an effective GHG forcing as follows

$$Fe\ (GHGs) = 1.12\ [Fa\ (CO_2) + 1.4\ Fa\ (CH_4)], \qquad (8)$$

where Fa is the adjusted forcing, i.e., the planetary energy imbalance due to the GHG change after the stratospheric temperature has time to adjust to the gas change. Fe, the effective forcing, accounts for variable efficacies of different climate forcings (Hansen et al., 2005). Formulas for Fa of each gas are given by Hansen et al. (2000). The factor 1.4 converts the adjusted forcing of $CH_4$ to its effective forcing, Fe, which is greater than Fa mainly because of the effect of $CH_4$ on tropospheric ozone and stratospheric water vapor (Hansen et al., 2005). The factor 1.12 approximates the forcing by $N_2O$ changes, which are not as well preserved in the ice cores but have a strong positive correlation with $CO_2$ and $CH_4$ changes (Spahni et al., 2005). The factor 1.12 is smaller than the 1.15 used by Hansen et al. (2007), consistent with estimates of the $N_2O$ forcing in the current GISS radiation code and IPCC (2007). Our LGM-Holocene GHG forcing (Fig. 5c) is ~3 W/m², moderately larger than the 2.8 W/m² estimated by IPCC (2007) because of our larger effective $CH_4$ forcing.

Climate forcing due to surface albedo change is a function mainly of sea level, which implicitly defines ice sheet size. Albedo change due to LGM-Holocene vegetation change, much of which is inherent with ice sheet area change, and albedo change due to coastline movement are lumped together with ice sheet area change in calculating surface albedo climate forcing. Ice sheet forcing does not depend sensitively on ice sheet shape or on how many ice sheets the ice volume is divided among and is nearly linear in sea level change (Fig. S4, Hansen et al., 2008). For simplicity, we use the linear relation in Hansen et al. (2008) Fig. S4, thus 5 W/m² between the LGM and ice-free conditions and 3.4 W/m² between the LGM and Holocene. This scale factor was based on simulations with an early climate model (Hansen et al., 1983, 1984); comparable forcings are found in other models (e.g., see discussion in Hewitt and Mitchell, 1997), but results depend on cloud representations, assumed ice albedo, and other factors, so the uncertainty is difficult to quantify. We subjectively estimate an uncertainty ~20 percent.

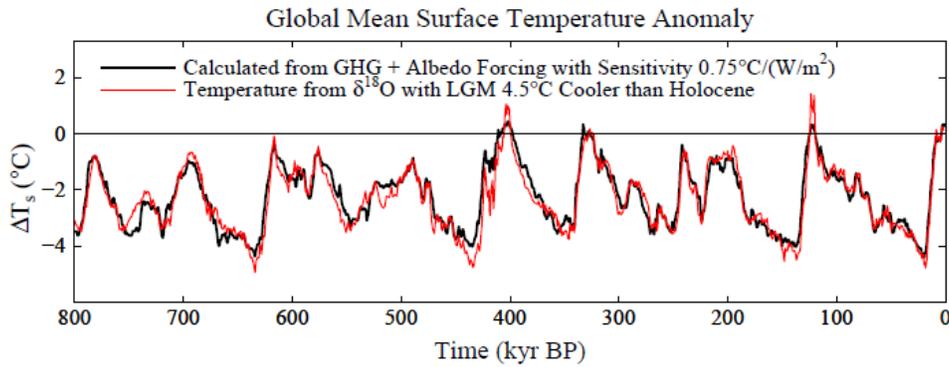

**Fig. 6.** Calculated surface temperature for forcings of Fig. 5c with climate sensitivity 0.75 °C per W/m², compared with 2×ΔTdo. Zero point is Holocene (10 kyr) mean.

Global temperature change obtained by multiplying the sum of the two climate forcings in Fig. 5c by sensitivity ¾°C per W/m² yields a remarkably good fit to "observations" (Fig. 6), where observed temperature is 2×ΔT$_{do}$, with 2 being the scale factor needed to yield the estimated 4.5°C LGM-Holocene surface temperature change. The close match is partly a result of the fact that sea level and temperature data are derived from the same deep ocean record, but use of other sea level reconstructions still yields a good fit between calculated and observed temperature (Hansen et al., 2008). However, exactly the same match as in Fig. 6 is achieved with fast-feedback sensitivity 1°C per W/m² if LGM cooling is 6°C or with sensitivity 0.5°C per W/m² if LGM cooling is 3°C.

Accurate data defining LGM-Holocene warming would aid empirical evaluation of fast-feedback climate sensitivity. Remarkably, the range of recent estimates of LGM-Holocene warming, from ~3°C (Schmittner et al., 2011) to ~6°C (Schneider von Deimling et al., 2006), is about the same as at the time of the CLIMAP (1981) project. Given today's much improved analytic capabilities, a new project to define LGM climate conditions, analogous to the PRISM Pliocene data reconstruction (Dowsett et al., 2009, 2010) and PlioMIP model intercomparisons (Haywood et al., 2010, 2012), could be beneficial. In section 7b we suggest that a study of Eemian glacial-interglacial climate change could be even more definitive. Combined LGM, Eemian and Pliocene studies would address an issue raised at a recent workshop (PALAEOSENS Project Members, 2012): the need to evaluate how climate sensitivity varies as a function of the initial climate state. The calculations below were initiated after the workshop as another way to address that question.

*(b) Fast-Feedback Sensitivity: State Dependence*

Climate sensitivity must be a strong function of the climate state. Simple climate models show that when Earth becomes cold enough for ice cover to approach the tropics, the amplifying albedo feedback causes rapid ice growth to the equator, "snowball Earth" conditions (Budyko, 1969). Real world complexity, including ocean dynamics, can mute this sharp bifurcation to a temporarily stable state (Pierrehumbert et al., 2011), but snowball events have occurred several times in Earth's history when the younger Sun was dimmer than today (Kirschvink, 1992). Earth escaped snowball conditions due to limited weathering in that state, which allowed volcanic $CO_2$ to accumulate in the atmosphere until there was enough $CO_2$ for the high sensitivity to cause rapid deglaciation (Hoffman and Schrag, 2002).

Climate sensitivity at the other extreme, as Earth becomes hotter, is also driven mainly by an $H_2O$ feedback. As climate forcing and temperature increase, the amount of water vapor in the air increases and clouds may change. Increased water vapor makes the atmosphere more opaque in the infrared region that radiates Earth's heat to space, causing the radiation to emerge from higher colder layers, thus reducing the energy emitted to space. This amplifying feedback has been known for centuries and was described remarkably well by Tyndall (1861). Ingersoll (1969) discussed water vapor's role in the "runaway greenhouse effect" that caused the surface of Venus to eventually become so hot that carbon was "baked" from the planet's crust, creating a hothouse climate with almost 100 bars of $CO_2$ in the air and surface temperature about 450°C, a stable state from which there is no escape. Arrival at this terminal state required passing through a "moist greenhouse" state in which surface water evaporates, water vapor becomes a major constituent of the atmosphere, and $H_2O$ is dissociated in the upper atmosphere with the hydrogen slowly escaping to space (Kasting, 1988). That Venus had a primordial ocean, with most of the water subsequently lost to space, is confirmed by present enrichment of deuterium over ordinary hydrogen by a factor of 100 (Donahue et al., 1982), the heavier deuterium being less efficient in escaping gravity to space.

The physics that must be included to investigate the moist greenhouse is principally: (1) accurate radiation incorporating the spectral variation of gaseous absorption in both the solar radiation and thermal emission spectral regions, (2) atmospheric dynamics and convection with no specifications favoring artificial atmospheric boundaries, such as between a troposphere and stratosphere, (3) realistic water vapor physics, including its effect on atmospheric mass and surface pressure, (4) cloud properties that respond realistically to climate change. Conventional global climate models are inappropriate, as they contain too much other detail in the form of parameterizations or approximations that break down as climate conditions become extreme.

We employ the simplified atmosphere-ocean model of Russell et al. (1995), which solves the same fundamental equations (conservation of energy, momentum, mass, and water substance, and the ideal gas law) as in more elaborate global models. Principal changes in the model physics in the current version of the model are use of a step-mountain C-grid atmospheric vertical-coordinate (Russell, 2007), addition of a drag in the grid-scale momentum equation in both atmosphere and ocean based on sub-grid topography variations, and inclusion of realistic ocean tides based on exact positioning of the Moon and Sun. Radiation is the k-distribution method of Lacis and Oinas (1991) with 25 k-values; the sensitivity of this specific radiation code is documented in detail by Hansen et al. (1997). Atmosphere and ocean dynamics are calculated on a 3°×4° Arakawa C-grids. There are 24 atmospheric layers. In our present simulations the ocean's depth is reduced to 100 m with 5 layers so as to achieve rapid equilibrium response to forcings; this depth limitation reduces poleward ocean transport by more than half. Moist convection is based on a test of moist static stability as in Hansen et al. (1983). Two cloud types occur: moist convective clouds, when the atmosphere is moist statically unstable, and large scale super-saturation, with cloud optical properties based on the amount of moisture removed to eliminate super-saturation, with scaling coefficients chosen to optimize the control run's fit with global observations (Russell et al., 1995 and paper in preparation). To avoid long response times in extreme climates, today's ice sheets are assigned surface properties of tundra, thus allowing them to have high albedo snow cover in cold climates but darker vegetation in warm climates. The model, the present experiments, and more extensive experiments will be described in a paper (Russell et al.) in preparation.

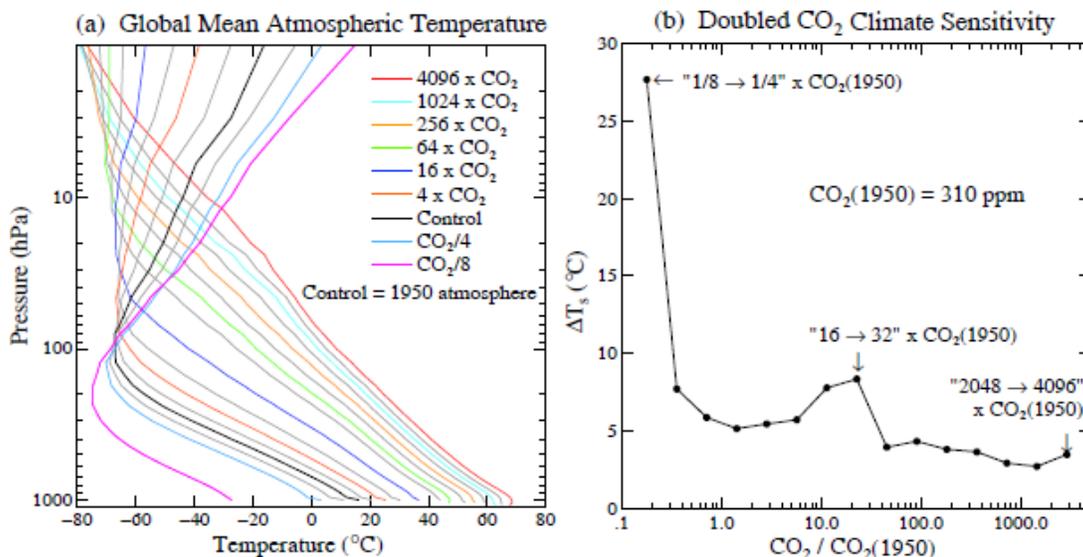

**Fig. 7.** (a) Calculated global mean temperature for successive doublings of $CO_2$ (legend identifies every other case), and (b) resulting climate sensitivity (1× $CO_2$ = 310 ppm).

Equilibrium response of the control run (1950 atmospheric composition, $CO_2$ ~ 310 ppm) and runs with successive $CO_2$ doublings and halvings reveal that the snowball Earth instability occurs just beyond 3 $CO_2$ halvings. Given that a $CO_2$ doubling or halving is equivalent to a 2% change of solar irradiance (Hansen et al., 2005) and the estimate that solar irradiance was ~6% lower 600 Myr ago at the most recent snowball Earth occurrence (Hoffman & Li, 2009), Fig. 7 implies that about 300 ppm $CO_2$ or less was sufficiently small to initiate glaciation at that time.

Climate sensitivity reaches large values at 8-32×$CO_2$ (~2500 to 10,000 ppm) (Fig. 7b). High sensitivity is caused by increasing water vapor as the tropopause rises and diminishing low cloud cover, but the sensitivity decreases for still larger $CO_2$ as cloud optical thickness and planetary albedo increase, as shown by Russell et al. (in preparation). The high sensitivity for $CO_2$ less than 4×$CO_2$ is due partly to the nature of the experiments (Greenland and Antarctic ice sheets being replaced by tundra). High albedo snow cover on these continents largely disappears between 1×$CO_2$ and 4×$CO_2$, thus elevating the calculated fast-feedback sensitivity from ~4°C to ~5°C. In the real world we would expect the Greenland and Antarctic ice sheets to be nearly eliminated and replaced by partially vegetated surfaces already at 2×$CO_2$ (620 ppm) equilibrium climate. In other words, if the Greenland/Antarctic surface albedo change were identified as a slow feedback, rather than as a fast-feedback snow effect as it is in Fig. 7, the fast-feedback sensitivity at 1-4×$CO_2$ would be ~4°C. Thus the sensitivity ~8°C per $CO_2$ doubling in the range 8-32×$CO_2$ is a very large increase over sensitivity at smaller $CO_2$ amounts.

How confident are we in the modeled fast-feedback sensitivity (Fig. 7b)? We suspect that the modeled water vapor feedback may be moderately exaggerated, because water vapor amount in the control run exceeds observed amounts. Also the area of sea ice in the control run exceeds observations, which may increase the modeled sensitivity in the 1-4×$CO_2$ range. On the other hand, we probably underestimate the sensitivity at very high $CO_2$ amounts, because our model (like most climate models) does not change total atmospheric mass as $CO_2$ amount varies. Mass change due to conceivable fossil fuel loading (up to say 16×$CO_2$) is unlikely to have much effect, but sensitivity is probably underestimated at high $CO_2$ amounts due to self-broadening of $CO_2$ absorption lines. The increased atmospheric mass is also likely to alter the cloud feedback, which otherwise is a strongly diminishing feedback at very large $CO_2$ amounts. Atmospheric mass will be important after Earth has lost its ocean and carbon is baked into the atmosphere. These issues are being examined in the Russell et al. paper in preparation.

Earth today, with ~1.26 times 1950 $CO_2$, is far removed from the "snowball" state. Because of the increase of solar irradiance over the past 600 million years and volcanic emissions, no feasible $CO_2$ amount could take Earth back to snowball conditions. Similarly, a Venus-like baked-crust $CO_2$ hothouse is far distant because it cannot occur until the ocean escapes to space. We calculate an escape time of order $10^8$-$10^9$ years even with the increased stratospheric water vapor and temperature at 16×$CO_2$. Given the transient nature of a fossil fuel $CO_2$ injection, the continuing forcing needed to achieve a terminal Venus-like baked-crust $CO_2$ hothouse must wait until the Sun's brightness has increased on the billion year time scale. However, the planet could become uninhabitable long before that.

The practical concern for humanity is the high climate sensitivity and the eventual climate response that may be reached if all fossil fuels are burned. Estimates of the carbon content of all fossil fuel reservoirs including unconventional fossil fuels such as tar sands, tar shale, and various gas reservoirs that can be tapped with developing technology (GEA, 2012) imply that $CO_2$ conceivably could reach a level as high as 16 times the 1950 atmospheric amount. In that event, Fig. 7 suggests a global mean warming approaching 25°C, with much larger warming at high latitudes (Fig. S6). The result would be a planet on which humans could work and survive outdoors in the summer only in mountainous regions (Kenney et al., 2004; Hanna and Brown, 1983) -- and there they would need to contend with the fact that a moist stratosphere would have destroyed the ozone layer (Kasting and Donahue, 1980).

## 6. Earth System Sensitivity

GHG and surface albedo changes, which we treated as specified climate forcings in evaluating fast-feedback climate sensitivity, are actually slow climate feedbacks during orbit-instigated Pleistocene glacial-interglacial climate swings. Given that GHG and albedo feedbacks are both strong amplifying feedbacks, indeed accounting by themselves for most of the global Pleistocene climate variation, it is apparent that today's climate sensitivity on millennial time scales must be substantially larger than the fast-feedback sensitivity.

Climate sensitivity including slow feedbacks is described as "Earth system sensitivity" (Lunt et al., 2010b; Pagani et al., 2010; Royer et al., 2012). There are alternative choices for the feedbacks included in Earth system sensitivity. Hansen and Sato (2012) suggest adding slow feedbacks one-by-one, creating a series of increasingly comprehensive Earth system climate sensitivities; specifically, they successively move climate-driven changes of surface albedo, non-$CO_2$ GHGs, and $CO_2$ into the feedback category, at which point the Earth system sensitivity is relevant to an external forcing such as changing solar irradiance or human-made forcings. At each level in this series the sensitivity is state-dependent.

Our principal aim here is to use Cenozoic climate change to infer information on the all-important fast-feedback climate sensitivity, including its state dependence, via analysis of Earth system sensitivity. $CO_2$ is clearly the dominant forcing of the long-term Cenozoic cooling, in view of the abundant evidence that $CO_2$ reached levels of the order of 1000 ppm in the early Cenozoic (Beerling and Royer, 2011), as discussed in the Overview above. Thus our approach is to examine Earth system sensitivity to $CO_2$ change by calculating the $CO_2$ history required to produce our reconstructed Cenozoic temperature history for alternative state-independent and state-dependent climate sensitivities. By comparing the resulting $CO_2$ histories with $CO_2$ proxy data we thus assess the most realistic range for climate sensitivity.

Two principal uncertainties in this analysis are (1) global temperature at the Early Eocene Climatic Optimum (EECO) ~ 50 Myr BP, and (2) $CO_2$ amount at that time. We use EECO ~ 28°C (Fig. 4) as our standard case, but we repeat the analysis with EECO ~ 33°C (Fig. S3), thus allowing inference of how the conclusions change if knowledge of Eocene temperature changes.

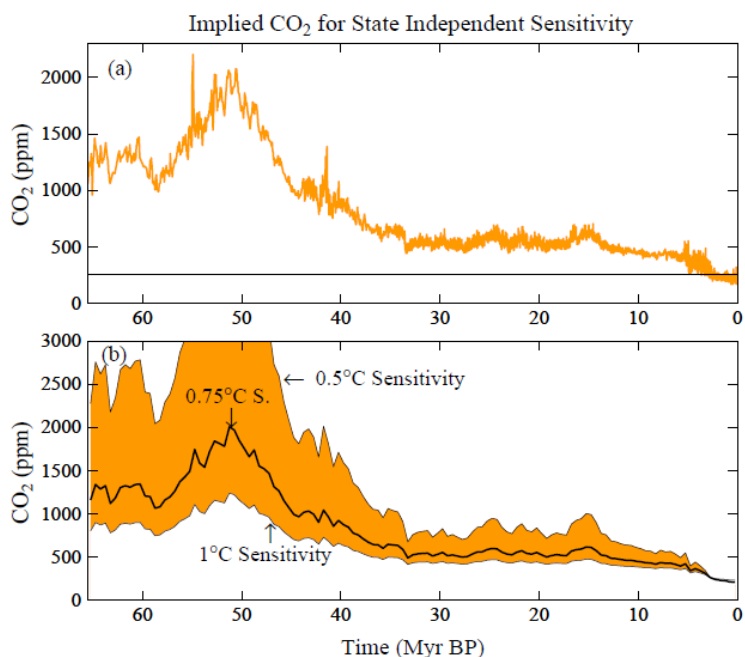

**Fig. 8.** (a) $CO_2$ amount required to yield global temperature of Fig. 4a if fast-feedback climate sensitivity is 0.75°C per W/m$^2$ and non-$CO_2$ GHGs contribute 25% of the GHG forcing, (b) same as in (a), but with temporal resolution 0.5 Myr and for three choices of fast-feedback sensitivity; $CO_2$ peak exceeds 5000 ppm in the case of 0.5°C sensitivity. Horizontal line is the early-mid Holocene 260 ppm $CO_2$ level.

Similarly, our graphs allow the inferred climate sensitivity to be adjusted if improved knowledge of $CO_2$ 50 Myr BP indicates a value significantly different than ~1000 ppm.

To clarify our calculations, let us first assume that fast-feedback climate sensitivity is a constant (state-independent) 3°C for doubled $CO_2$ (0.75°C per W/m$^2$). It is then trivial to convert our global temperature for the Cenozoic (Fig. 4a) to the total climate forcing throughout the Cenozoic, which is shown in Fig. S4a, as are results of subsequent steps. Next we subtract the solar forcing, a linear increase of 1 W/m$^2$ over the Cenozoic era due to the Sun's 0.4% irradiance increase (Sackmann et al., 1993), and the surface albedo forcing due to changing ice sheet size, which we take as linear at 5 W/m$^2$ for 180 m sea level change from 35 Myr BP to the LGM. These top-of-the-atmosphere and surface forcings are moderate in size, compared to the total forcing over the Cenozoic, and partially offsetting, as shown in Fig. S4b. The residual forcing, which has a maximum ~17 W/m$^2$ just prior to 50 Myr BP, is the atmospheric forcing due to GHGs. Non-$CO_2$ GHGs contribute 25% of the total GHG forcing in the period of ice core measurements. Atmospheric chemistry simulations (Beerling et al., 2011) reveal continued growth of non-$CO_2$ gases ($N_2O$, $CH_4$ and tropospheric $O_3$) in warmer climates, at only a slightly lower rate (1.7-2.3 W/m$^2$ for 4×$CO_2$, which itself is ~8 W/m$^2$ ). Thus we take the $CO_2$ forcing as 75% of the GHG forcing throughout the Cenozoic in our standard case, but we also consider the extreme case in which non-$CO_2$ gases are fixed and thus contribute no climate forcing.

$CO_2$ forcing is readily converted to $CO_2$ amount; we use the equation in Table 1 of Hansen et al. (2000). The resulting Cenozoic $CO_2$ history required to yield the global surface temperature of Fig. 4a is shown in Fig. 8a for state-independent climate sensitivity with non-$CO_2$ GHGs providing 25% of the GHG climate forcing. Peak $CO_2$ in this case is ~2000 ppm. If non-$CO_2$ GHGs provide less than 25% of the total GHG forcing, the inferred $CO_2$ amount would be even greater. Results for alternative sensitivities, as in Fig. 8b, are calculated for temporal resolution 0.5 Myr to smooth out glacial-interglacial $CO_2$ oscillations, as our interest here is in $CO_2$ as a climate forcing.

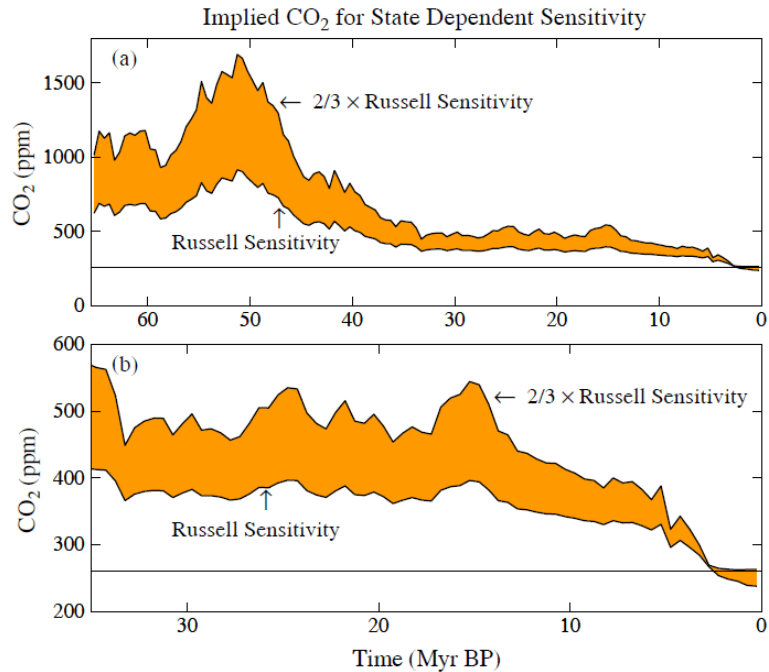

**Fig. 9.** (a) $CO_2$ amount required to yield global temperature history of Fig. 4a if fast-feedback climate sensitivity is that calculated with the Russell model, i.e., the sensitivity shown in Fig. 7b, and two-thirds of that sensitivity. These results assume that non-$CO_2$ GHGs provide 25% of the GHG climate forcing. (b) vertical expansion for past 35 Myr.

We focus on the $CO_2$ amount 50 Myr BP averaged over a few million years in assessing the realism of our inferred $CO_2$ histories, because $CO_2$ variations in the Cenozoic remain very uncertain despite the success of Beerling and Royer (2011) in eliminating the most extreme outliers. Beerling and Royer (2011) find a best-fit $CO_2$ at 50 Myr BP of about 1000 ppm -- see their Fig. 1, which also indicates that $CO_2$ at 50 Myr BP was almost certainly in the range 750-1500 ppm, even though it is impossible to provide a rigorous confidence interval.

We conclude that the average fast-feedback climate sensitivity during the Cenozoic is larger than the canonical 3°C for 2×$CO_2$ (0.75°C per W/m$^2$) that has long been the central estimate for current climate. An average 4°C for 2×$CO_2$ (1°C per W/m$^2$) provides a good fit to the target 1000 ppm $CO_2$, but the sensitivity must be still higher if non-$CO_2$ GHG forcings amplify the $CO_2$ by less than one-third, i.e., provide less than 25% of total GHG forcing.

*(a) State-Dependent Climate Sensitivity*

More realistic assessment should account for the state-dependence of climate sensitivity. Thus we make the same calculations for the state-dependent climate sensitivity of the Russell climate model, i.e., we use the fast-feedback climate sensitivity of Fig. 7b. In addition, for the purpose of assessing how the results depend upon climate sensitivity, we consider a second case in which we reduce the Russell sensitivity of Fig. 7b by the factor 2/3.

The estimated 1000 ppm of $CO_2$ at 50 Myr BP falls between the Russell sensitivity and two-thirds of the Russell sensitivity, though closer to the full Russell sensitivity. If the non-$CO_2$ GHG forcing is less than one-third of the $CO_2$ forcing, the result is even closer to the full Russell sensitivity. With these comparisons at 50 Myr BP in mind, we can use Fig. 9 to infer the likely $CO_2$ amount at other times. The end-Eocene transition began at about 500 ppm and fell to about 400 ppm. The mid-Miocene warmth, which peaked at about 15 Myr BP, required a $CO_2$ increase of only a few tens of ppm with the Russell sensitivity, but closer to 100 ppm if the true sensitivity is only two-thirds of the Russell sensitivity. The higher (full Russell) sensitivity requires much less $CO_2$ change to produce the mid-Miocene warming for two reasons: (1) the greater temperature change for a specified forcing, and (2) the smaller $CO_2$ change required to yield a given forcing from the lesser $CO_2$ level of the higher sensitivity case. Average $CO_2$ amount in the early Pliocene is about 300 ppm for the Russell sensitivity, but could reach a few tens of ppm higher if the true sensitivity is closer to two-thirds of the Russell sensitivity.

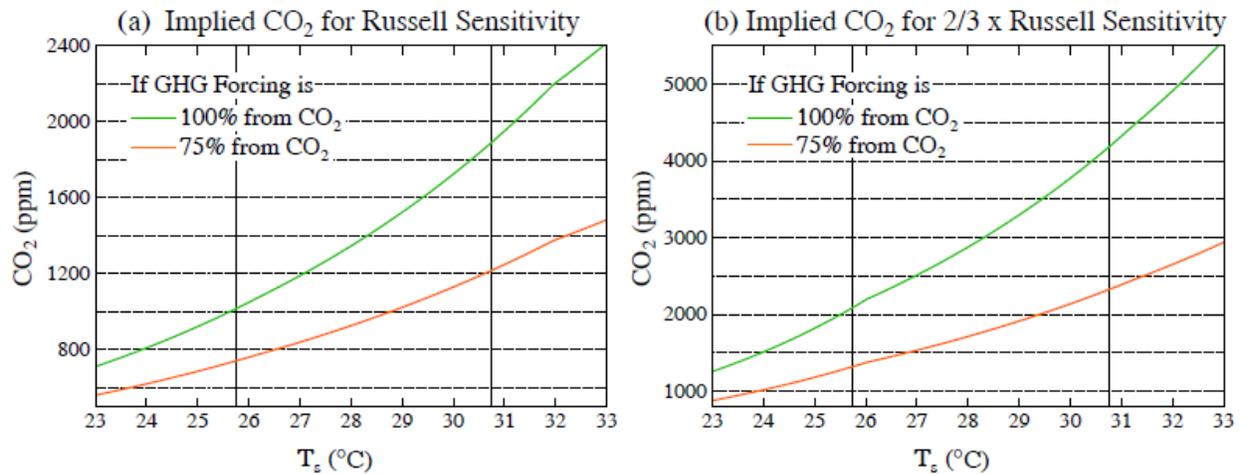

**Fig. 10.** Atmospheric $CO_2$ amount (y-axis) required to yield a given global temperature (x-axis) at the time of the PETM for (a) the Russell climate sensitivity, and (b) two-thirds of the Russell sensitivity. The $CO_2$ increment required to yield a given PETM warming above the pre-PETM temperature (25.7°C) is obtained by subtracting the $CO_2$ amount at the desired Ts from the $CO_2$ at Ts = 25.7°C. The vertical line is for the case of 5°C PETM warming. The orange lines show the required $CO_2$ if the $CO_2$ increase is accompanied by a non-$CO_2$ GHG feedback that provides 25 percent of the total GHG forcing.

*(b) Comparison to van de Wal et al. Model*

van de Wal et al. (2011) used the same Zachos et al. (2008) $\delta^{18}O$ data to drive an inverse model calculation, including an ice sheet model to separate ice volume and temperature, thus inferring $CO_2$ over the past 20 Myr. They find a Mid-Miocene Climatic Optimum (MMCO) $CO_2$ ~450 ppm, which falls between the Russell and 2/3 Russell sensitivities (Fig. 9). The van de Wal et al. (2011) model has 30°C change of Northern Hemisphere temperature (their model is hemispheric) between the MMCO and average Pleistocene conditions driven by a $CO_2$ decline from ~450 ppm to ~250 ppm, which is a forcing ~3.5 W/m². Thus the implied (Northern Hemisphere) Earth system sensitivity is an implausible ~35°C for a 4 W/m² $CO_2$ forcing. The large temperature change may be required to produce substantial sea level change in their ice sheet model, which we suggested above is unrealistically unresponsive to climate change. However, they assign most of the temperature change to slow feedbacks, thus inferring a fast-feedback sensitivity of only about 3°C per $CO_2$ doubling.

*c) Inferences from the PETM and Early Cenozoic Climate*

Finally, we use the largest and best documented of the hyperthermals, the Paleocene Eocene Thermal Maximum (PETM) to test the reasonableness of the Russell state-dependent climate sensitivity. Global warming in the PETM is reasonably well-defined at 5-6°C and the plausible range for carbon mass input is approximately 4000-7000 PgC (Dunkley Jones et al., 2010). Given that the PETM carbon injection occurred over a period of a few millennia, carbon cycle models suggest that about one-third of the carbon would be airborne as $CO_2$ following complete injection (Archer, 2005). With conversion factor 1 ppm $CO_2$ ~ 2.12 GtC, the 4000-7000 GtC source thus yields ~630-1100 ppm $CO_2$. We can use Fig. 10, obtained via the same calculations as described above, to see how much $CO_2$ is required to yield 5°C warming. The Russell sensitivity requires ~800 ppm $CO_2$ for 5°C warming, while 2/3 of the Russell sensitivity requires ~2100 ppm $CO_2$. Given uncertainty in the airborne fraction of $CO_2$ and possible non-$CO_2$ gases, we cannot rule out the 2/3 Russell sensitivity, but the full Russell sensitivity fits plausible PETM carbon sources much better, especially if the PETM warming is actually somewhat more than 5°C (see Fig. 10 for quantitative implications).

This analysis is for Earth system sensitivity with $CO_2$ as the forcing, as is appropriate for the PETM because any carbon injected as $CH_4$ would be rapidly oxidized to $CO_2$. Feedbacks in the PETM do not include large ice sheets, but non-$CO_2$ GHGs are an unmeasured feedback. If a warming climate increases the amount of $N_2O$ and $CH_4$ in the air, the required carbon source for a given global warming is reduced, because the amount of carbon in airborne $CH_4$ is negligible. Any non-$CO_2$ GHG feedback increases the $CO_2$-forced Earth system sensitivity, potentially by a large amount (see Fig. 10). The $CO_2$-forced Earth sensitivity is the most relevant climate sensitivity, not only for the PETM but for human-made forcings. Although present enhanced amounts of airborne $CH_4$ and $N_2O$ are mostly a climate forcing, i.e., their increases above the preindustrial level are mainly a consequence of human-made sources, they also include a GHG feedback. Climate sensitivity including this GHG feedback is the most relevant sensitivity for humanity as the $CO_2$ forcing continues to grow.

If EECO (Early Eocene Climatic Optimum) global temperature exceeded 28°C, as suggested by multi-proxy data taken at face value (see above), climate sensitivity implied by the EECO warmth and the PETM warming is close to the full Russell climate sensitivity (see Figs. S7, S8 and S9). We conclude that the existing data favor a climate sensitivity of at least 2/3 of the Russell sensitivity, and likely closer to the full Russell sensitivity. That lower limit is just over 3°C for 2×$CO_2$ for the range of climate states of immediate relevance to humanity (Fig. 7b).

## 7. Summary Discussion

Co-variation of climate, sea level, and atmospheric $CO_2$ through the Cenozoic era is a rich source of information that can advise us about the sensitivity of climate and ice sheets to forcings, including human-made forcings. Our approach is to estimate Cenozoic sea level and temperature from empirical data, with transparent assumptions and minimal modeling. Our data are available in the Supporting Material, allowing comparison with other data and model results.

*(a) Sea Level Sensitivity*

Hansen (2005, 2007) argues that real ice sheets are more responsive to warming than in most ice sheet models, which suggest that large ice sheets are relatively stable. The model of Pollard and DeConto (2005), for example, requires 3-4 times the pre-industrial $CO_2$ amount to melt the Antarctic ice sheet. This stability is in part a result of hysteresis: as Earth warms, the ice sheet size as a function of temperature does not return on the same path that it followed as temperature fell and the ice sheet grew. We do not question the reality of mechanisms that cause ice sheet hysteresis, but we suspect they are exaggerated in models. Thus, as an extreme alternative that can be compared with ice sheet models and real world data, we assume that hysteresis effects are negligible in our approximation for sea level as a function of temperature.

Ice sheets in question are those on Greenland and Antarctica, ice sheets that could shrink with future warming. Despite stability of those ice sheets in the Holocene, there is evidence that sea level was much more variable during the Eemian, when we estimate peak global temperature was only +1.0°C warmer than in the first decade of the 21st century. Rohling et al. (2008) estimate an average rate of Eemian sea level change of 1.4 m/century, and several studies noted above suggest that Eemian sea level reached heights of +4-6 m or more relative to today.

The Mid-Miocene Climatic Optimum provides one test of hysteresis. Our sea level approximation (Fig. 2) suggests that the Antarctic ice sheet nearly disappeared at that time. John et al. (2011) provide support for that interpretation, as well as evidence of numerous rises and falls of sea level by 20-30 m during the Miocene. These variations are even larger than those we find (Fig. 2), but the resolution of the $\delta^{18}O$ data we employ is not adequate to provide the full amplitude of variations during that period (Fig. S1).

The Mid-Pliocene is a more important test of ice sheet variability. We find sea level fluctuations of at least 20-40 m, much greater than in ice sheet models (Fig. 2), with global temperature variations of only a few degrees. Independent analyses designed to separate ice volume and temperature change, such as Dwyer and Chandler (2009), find sea level maxima and variability comparable to our estimates. Altogether, the empirical data support a high sensitivity of sea level to global temperature change, and they provide strong evidence against the seeming lethargy and large hysteresis effects that occur in at least some ice sheet models.

*(b) Fast-Feedback Climate Sensitivity*

Estimates of climate sensitivity cover a wide range that has existed for decades (IPCC, 2007; PALAEOSENS, 2012). That range measures our ignorance; it does not mean that climate response from a specified state is stochastic with such inherent uncertainty. God (Nature) plays dice, but not for such large amounts. Indeed, one implication of the tight fit of calculated and measured temperature change of the past 800,000 years (Fig. 6) is that there is a single well-defined, but unknown, fast-feedback global climate sensitivity for that range of climate, despite large regional climate variations and ocean dynamical effects (Masson-Delmotte et al., 2010).

Improved empirical data can define climate sensitivity much more precisely, provided that climate-induced aerosol changes are included in the category of fast feedbacks (human-made aerosol changes are a climate forcing). Empirical assessment of fast-feedback climate sensitivity is obtained by comparing two quasi-equilibrium climate states for which boundary condition climate forcings (which may be slow feedbacks) are known. Aerosol changes between those climate states are appropriately included as a fast feedback, not only because aerosols respond rapidly to changing climate but also because there are multiple aerosol compositions, they have complex radiative properties, and they affect clouds in several ways, thus making accurate knowledge of their glacial-interglacial changes inaccessible.

The temporal variation of the GHG plus surface albedo climate forcing closely mimics the temporal variation of either the deep ocean temperature (Fig. 6) or Antarctic temperature (Masson-Delmotte et al., 2010; Hansen et al., 2008) for the entire 800,000 years of polar ice core data. However, the temperature change must be converted to the global mean to allow inference of climate sensitivity. The required scale factor is commonly extracted from an estimated LGM-Holocene global temperature change, which, however, is very uncertain, with estimates ranging from ~3°C to ~6°C. Thus, for example, the climate sensitivity (1.7-2.6°C for 2×$CO_2$) estimated by Schmittner et al. (2011) is due largely to their assumed ~3°C cooling in the LGM, and in lesser part to the fact that they defined some aerosol changes (dust) to be a climate forcing.

Climate sensitivity extracted from Pleistocene climate change is thus inherently partly subjective as it depends on how much weight is given to mutually inconsistent estimates of glacial-to-interglacial global temperature change. Our initial assessment is a fast-feedback sensitivity of 3 ± 1°C for 2×$CO_2$, corresponding to an LGM cooling of 4.5°C, similar to the 2.2-4.8°C estimate of PALAEOSENS (2012). This sensitivity is higher than estimated by Schmittner et al. (2011) partly because they included natural aerosol changes as a forcing. Also we note that their proxies for LGM sea surface cooling exclude planktic foraminifera data, which suggest larger cooling (Nurnberg et al., 2000), and, as noted by Schneider von Deimling et al.

(2006), regions that are not sampled tend to be ones where the largest cooling is expected. It should be possible to gain consensus on a narrower range for climate sensitivity via a community project for the LGM analogous to PRISM Pliocene data reconstruction (Dowsett et al., 2009, 2010) and PlioMIP model intercomparisons (Haywood et al., 2010, 2012).

However, we suggest that an even more fruitful approach would be a focused effort to define the glacial-to-interglacial climate change of the Eemian period (MIS 5e). The Eemian avoids the possibility of significant human-made effects, which may be a factor in the Holocene. Ruddiman (2003) suggests that deforestation and agricultural activities affected $CO_2$ and $CH_4$ in the Holocene, and Hansen et al. (2007) argue that human-made aerosols likely were important. Given the level of Eemian warmth, ~ +1.8°C relative to 1880-1920, with climate forcing similar to that for LGM-Holocene (Fig. 5), we conclude that this relatively clean empirical assessment yields a fast-feedback climate sensitivity in the upper part of the range suggested by the LGM-Holocene climate change, i.e., a sensitivity 3-4°C for 2×$CO_2$. Detailed study is especially warranted because Eemian warmth is anticipated to reoccur in the near-term.

### (c) Earth System Sensitivity

We have shown that global temperature change over the Cenozoic era is consistent with $CO_2$ change being the climate forcing that drove the long-term climate change. Proxy $CO_2$ measurements are so variable and uncertain that we only rely on the conclusion that $CO_2$ amount was of the order of 1000 ppm during peak early Eocene warmth. That conclusion, in conjunction with a climate model incorporating only the most fundamental processes, constrains average fast-feedback climate sensitivity to be in the upper part of the sensitivity range that is normally quoted (IPCC, 2007; PALAEOSENS, 2012), i.e., the sensitivity is greater than 3°C for 2×$CO_2$. Strictly this Cenozoic evaluation refers to the average fast-feedback sensitivity for the range of climates from ice ages to peak Cenozoic warmth and to the situation at the time of the PETM. However, it would be difficult to achieve that high average sensitivity if the current fast-feedback sensitivity were not at least in the upper half of the range $3 \pm 1$°C for 2×$CO_2$.

This climate sensitivity evaluation has implications for atmospheric $CO_2$ amount throughout the Cenozoic era, which can be checked as improved proxy $CO_2$ measurements become available. $CO_2$ amount was only ~450-500 ppm 34 Myr BP when large scale glaciation first occurred on Antarctica. Perhaps more important, the amount of $CO_2$ required to melt most of Antarctica in the Mid-Miocene Climatic Optimum was only ~450-500 ppm, conceivably only about 400 ppm. These $CO_2$ amounts are smaller than suggested by ice sheet/climate models, providing further indication that the ice sheet models are excessively lethargic, i.e., resistant to climate change. $CO_2$ amount in the earliest Pliocene, averaged over astronomical cycles, was apparently only about 300 ppm, and decreased further during the Pliocene.

### (d) Runaway Greenhouse

Our climate simulations, using a simplified three-dimensional climate model to solve the fundamental equations for conservation of water, atmospheric mass, energy, momentum, and the ideal gas law, but stripped to basic radiative, convective and dynamical processes, finds upturns in climate sensitivity at the same forcings as found with a more complex global climate model (Hansen et al., 2005). At forcings beyond these points the complex model "crashed", as have other climate models (discussed by Lunt et al., 2012). The upturn at 10-20 W/m$^2$ negative forcing has a simple physical explanation, it is the "snowball Earth" instability. Model crashes for large positive forcings are sometimes described as a runaway greenhouse, but they probably are caused by one of the many parameterizations in complex global models going outside its range of validity, not by a runaway greenhouse effect.

"Runaway greenhouse effect" has several meanings ranging from, at the low end, global warming sufficient to induce out-of-control amplifying feedbacks such as ice sheet disintegration and melting of methane hydrates, to, at the high end, a Venus-like hothouse with crustal carbon baked into the atmosphere and surface temperature of several hundred degrees, a climate state from which there is no escape. Between these extremes is the "moist greenhouse", which occurs if the climate forcing is large enough to make $H_2O$ a major atmospheric constituent (Kasting, 1988). In principle, an extreme moist greenhouse might cause an instability with water vapor preventing radiation to space of all absorbed solar energy, resulting in very high surface temperature and evaporation of the ocean (Ingersoll, 1969). However, the availability of non-radiative means for vertical transport of energy, including small-scale convection and large-scale atmospheric motions, must be accounted for, as is done in our atmospheric general circulation model. Our simulations indicate that no plausible human-made greenhouse gas forcing can cause an instability and runaway greenhouse effect as defined by Ingersoll (1969), in agreement with the theoretical analyses of Goldblatt and Watson (2012).

On the other hand, conceivable levels of human-made climate forcing could yield the low-end runaway greenhouse. A forcing of 12-16 $W/m^2$, which would require $CO_2$ increase by a factor of 8-16 times, if the forcing were due only to $CO_2$ change, would raise global mean temperature by 16-24°C with much larger polar warming. Surely that would melt all the ice on the planet, and likely thaw methane hydrates and scorch carbon from global peat deposits and tropical forests. This forcing would not produce the extreme Venus-like baked-crust greenhouse state, which cannot be reached until the ocean is lost to space. Warming of 16-24°C produces a moderately moist greenhouse, with water vapor increasing to about 1% of the atmosphere's mass, thus increasing the rate of hydrogen escape to space. However, if the forcing is by fossil fuel $CO_2$ the weathering process would remove the excess atmospheric $CO_2$ on a time scale of $10^4$-$10^5$ years, well before the ocean is significantly depleted. Baked-crust hot-house conditions on Earth require a large long-term forcing that is unlikely to occur until the sun brightens by a few tens of percent, which will take a few billion years (Sackmann et al., 1993).

*(e) Global Habitability*

Burning all fossil fuels would produce a different, practically uninhabitable, planet. Let us first consider a 12 $W/m^2$ greenhouse forcing, which we simulated with 8×$CO_2$. If non-$CO_2$ greenhouse gases such as $N_2O$ and $CH_4$ increase with global warming at the same rate as in the paleoclimate record and atmospheric chemistry simulations (Beerling et al., 2011), these other gases provide ~25 percent of the greenhouse forcing. The remaining 9 $W/m^2$ forcing requires ~4.8×$CO_2$, corresponding to fossil fuel emissions as much as ~10,000 GtC for a conservative assumption of a $CO_2$ airborne fraction averaging one-third over the 1000 years following peak emission (Archer, 2005; Archer et al., 2009).

Our calculated global warming in this case is 16°C, with warming at the poles about 30°C. Calculated warming over land areas averages ~20°C. Such temperatures would eliminate grain production in almost all agricultural regions in the world (Hatfield et al., 2011). Increased stratospheric water vapor would diminish the stratospheric ozone layer (Anderson et al., 2012).

More ominously, global warming of that magnitude would make much of the planet uninhabitable by humans (Sherwood and Huber, 2010; McMichael and Dear, 2010). The human body generates about 100 W of metabolic heat that must be carried away to maintain a core body temperature near 37°C, which implies that sustained wet bulb temperatures above 35°C can result in lethal hyperthermia (Sherwood and Huber, 2010; Dewhirst et al., 2003). Today summer temperature varies widely over Earth's surface, but wet bulb temperature is more narrowly confined by the effect of humidity, with the most common value ~26-27°C and highest ~31°C.

Warming of 10-12°C would put much of today's world population in regions with wet bulb temperature above 35°C (Sherwood and Huber, 2010). Given the 20°C warming we find with 4.8×$CO_2$, it is clear that such climate forcing would produce intolerable climatic conditions even if the true climate sensitivity is significantly less than the Russell sensitivity, or, if the Russell sensitivity is accurate, the $CO_2$ amount needed to produce intolerable conditions for humans is less than 4.8×$CO_2$. Note also that increased heat stress due to warming of the past few decades is already enough to affect health and workplace productivity at low latitudes, where the impact falls most heavily on low and middle-income countries (Kjellstrom et al., 2009).

Earth was 10-12°C warmer than today in the early Eocene and at the peak of the PETM (Fig. 4). How did mammals survive that warmth? Some mammals have higher internal temperatures than humans and there is evidence of evolution of surface-area-to-mass ratio to aid heat dissipation, e.g., transient dwarfing of mammals (Alroy et al., 2000) and even soil fauna (Smith et al., 2004) during the PETM warming. However, human-made warming will occur in a few centuries, as opposed to several millennia in the PETM, thus providing little opportunity for evolutionary dwarfism to alleviate impacts of global warming. We conclude that the large climate change from burning all fossil fuels would threaten the biological health and survival of humanity, making policies that rely substantially on adaptation inadequate.

Let us now verify that our assumed fossil fuel climate forcing of 9 W/m$^2$ is feasible. If we assume that fossil fuel emissions increase 3% per year, typical of the past decade and of the entire period since 1950, cumulative fossil fuel emissions will reach 10,000 GtC in 118 years. However, with such large rapidly growing emissions the assumed 33% $CO_2$ airborne fraction is surely too small. The airborne fraction, observed to have been 55% since 1950 (IPCC, 2007a), should increase because of well-known non-linearity in ocean chemistry and saturation of carbon sinks, implying that the airborne fraction probably will be closer to two-thirds rather than one-third, at least for a century or more. Thus the fossil fuel source required to yield 9 W/m$^2$ forcing may be closer to 5,000 GtC, rather than 10,000 GtC.

Are there sufficient fossil fuel reserves to yield 5,000-10,000 GtC? Recent updates of potential reserves (GEA, 2012), including unconventional fossil fuels (such as tar sands, tar shale, and hydrofracking-derived shale gas) in addition to conventional oil, gas and coal, suggest that 5×$CO_2$ (1400 ppm) is indeed feasible. For instance, using the emission factor for coal from IPCC (2007b), coal resources given by GEA (2012) amount to 7,300-11,000 GtC. Similarly, using emission factors from IPCC (2007b), total recoverable fossil energy reserves and resources estimated by GEA (2012) are ~15,000 GtC. This does not include large "additional occurrences" listed in Ch.7 of GEA (2012). Thus, for a multi-centennial $CO_2$ airborne fraction between one-third to two-thirds, as discussed above, there are more than enough available fossil fuels to cause a forcing of 9 W/m$^2$ sustained for centuries.

Most remaining fossil fuel carbon is in coal and unconventional oil and gas. Thus, it seems, humanity stands at a fork in the road. As conventional oil and gas are depleted, will we move to carbon-free energy and efficiency -- or to unconventional fossil fuels and coal? If fossil fuels were made to pay their costs to society, costs of pollution and climate change, carbon-free alternatives might supplant fossil fuels over a period of decades. However, if governments force the public to bear the external costs and even subsidize fossil fuels, carbon emissions are likely to continue to grow, with deleterious consequences for young people and future generations.

It seems implausible that humanity will not alter its energy course as consequences of burning all fossil fuels become clearer Yet strong evidence about the dangers of human-made climate change have so far had little effect. Whether governments continue to be so foolhardy as to allow or encourage development of all fossil fuels may determine the fate of humanity.

**Acknowledgments.** We thank James Zachos for the deep ocean oxygen isotope data, Chris Brierly, Mark Chandler, Bas de Boer, Alexey Fedorov, Chris Hatfield, Dorothy Peteet, David Rind, Robert Rohde and Cynthia Rosenzweig for helpful information, Andy Ridgwell for useful editorial suggestions and patience, Eelco Rohling for ably organizing the paleoclimate workshop that spurred writing of our present paper, Gerry Lenfest (Lenfest Foundation), ClimateWorks, Lee Wasserman (Rockefeller Family Foundation), Stephen Toben (Flora Family Foundation) and NASA program managers Jack Kaye and David Considine for research support.

**Supplementary Material**

Purposes of our Supplementary Material are to clarify characteristics of the data we employ in our analysis, provide additional comparisons with other analyses, clarify our calculations of Cenozoic climate forcings and temperature, and make the data in our figures available in digital form.

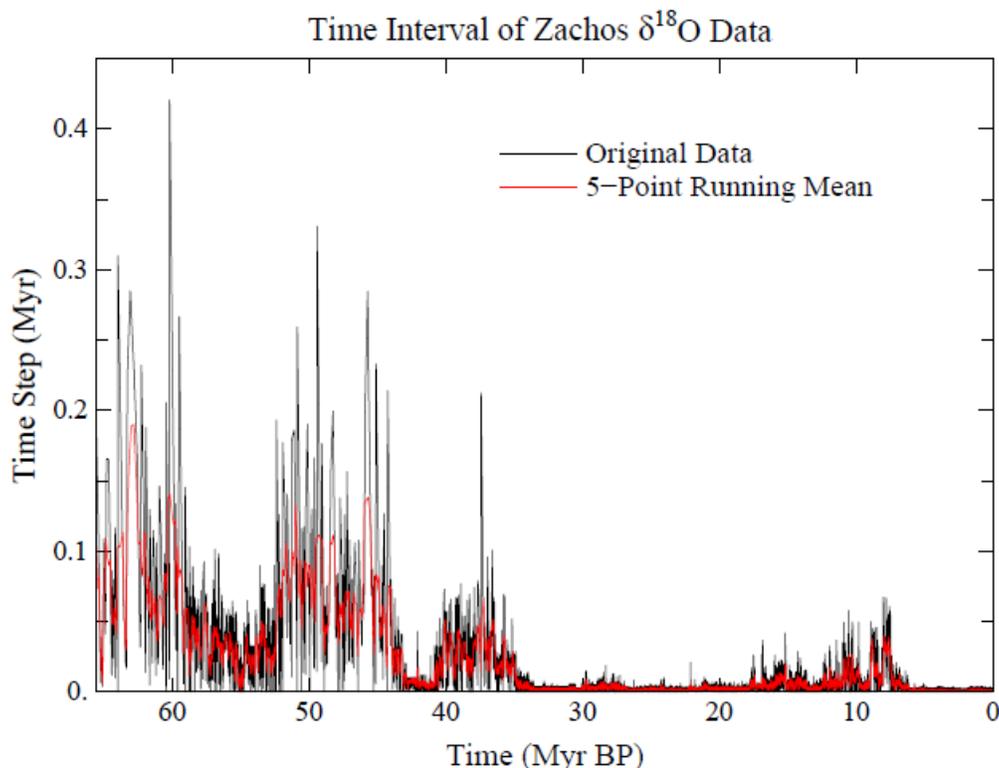

**Fig. S1.** This graph shows the time interval between successive data points in the Zachos et al. (2008) $\delta^{18}O$ data set and the 5-point running mean of that time interval.

The coarser temporal resolution in the early half of the Cenozoic and between 16 and 7 Myr BP has an effect on how variable the inferred temperature and sea level appear to be in our reconstructions (in the figures in the paper and in the Supplementary Material). Specifically, the apparent magnitude of glacial-interglacial oscillations of sea level and temperature is reduced in the periods with coarse temporal resolution.

A table (less than 1 Mb) with the $\delta^{18}O$ data at original resolution and our calculated 5-point running means of $\delta^{18}O$, deep ocean temperature, surface temperature, and sea level is available from the web site http://www.columbia.edu/~mhs119/ with this specific table located at http://www.columbia.edu/~mhs119/Sensitivity+SL+CO2/Table.txt

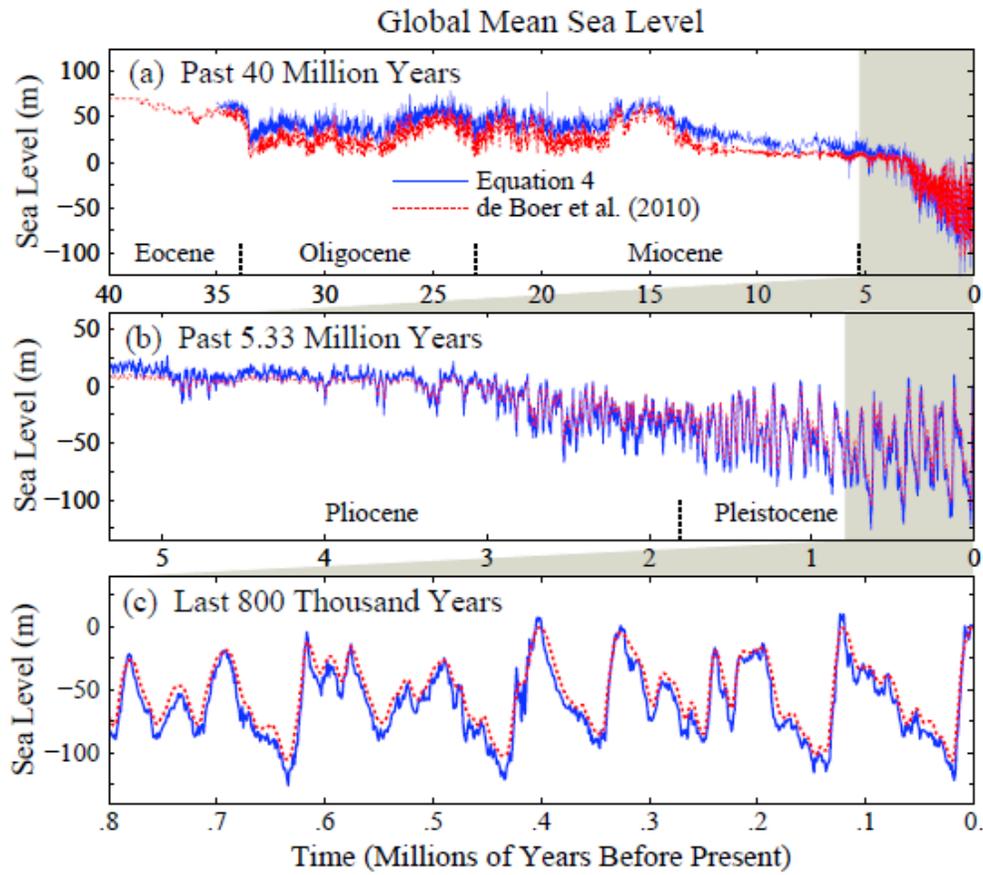

**Fig. S2.** Comparison of sea level from ice sheet model of de Boer et al. (2110) with reconstructions based on our equations (3) and (4) using $\delta^{18}O$ data of Zachos et al. (2008).

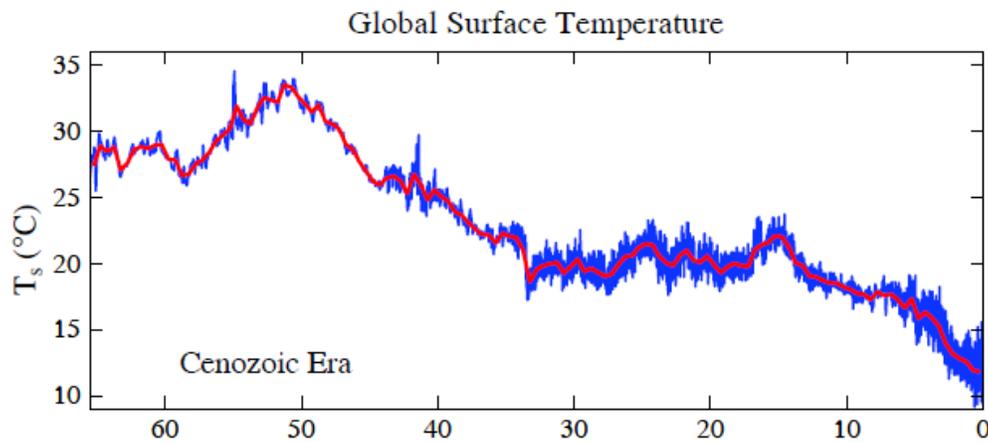

**Fig. S3.** Global surface temperature if $\Delta T_s = 1.5 \times \Delta T_{do}$ prior to 5.33 Myr BP, otherwise same as Fig. 4.

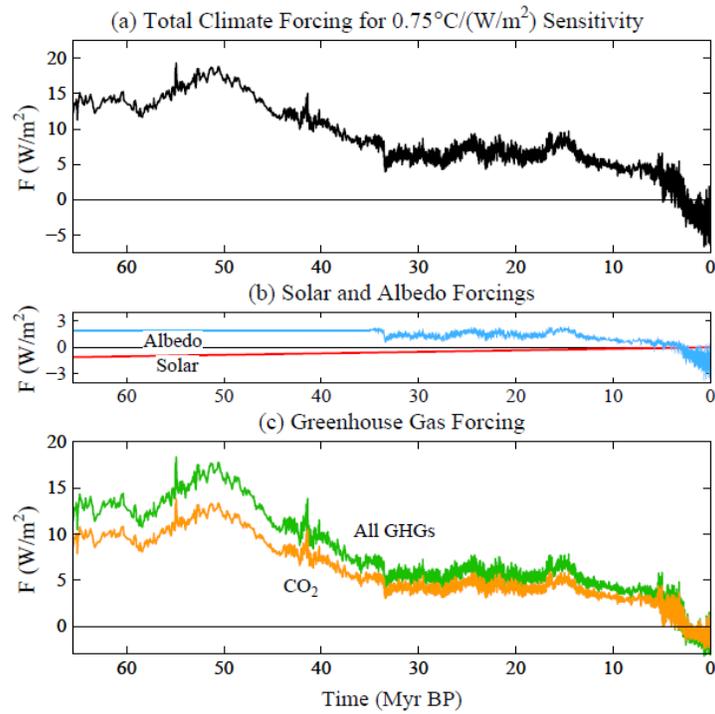

**Fig. S4.** (a) Climate forcing required to yield surface temperature of our Fig. 4 if climate sensitivity is a constant (state-independent) 3°C for 2×$CO_2$ (0.75°C per W/m$^2$). (b) Solar and albedo forcings that are next subtracted from the total forcing of (a) to obtain (c) the atmospheric forcing required to yield the surface temperature history in Fig. 4. The $CO_2$ forcing illustrated here is based on the assumption that $CO_2$ forcing is 75% of the total GHG forcing.

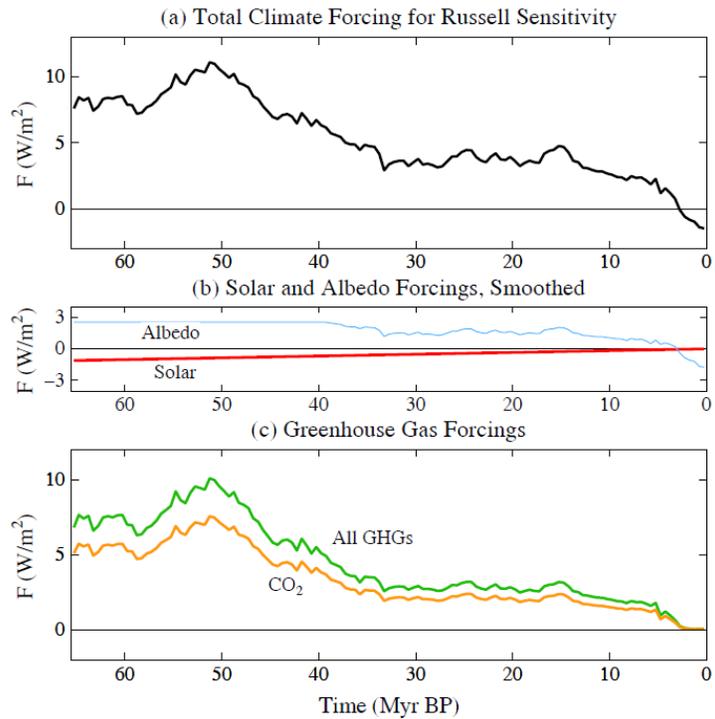

**Fig. S5.** As in Fig. S4, but the climate sensitivity is the state-dependent Russell sensitivity and the temperature record is the 500 kyr resolution curve in Fig. 4.

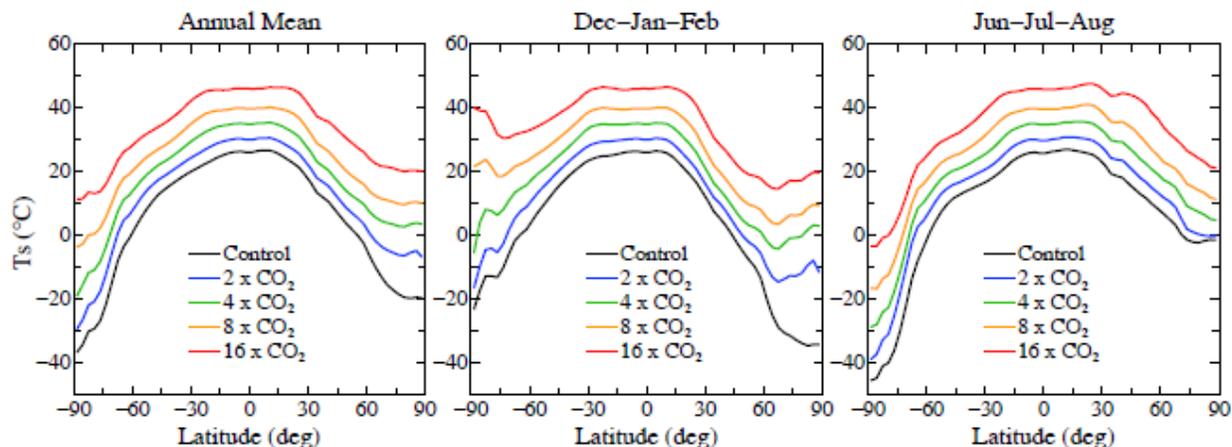

**Fig. S6.** Surface temperature versus latitude in simulations with the Russell climate model with a 100 m ocean. the control run has 310 ppm $CO_2$. Non-$CO_2$ gases were fixed, so simulations had 620, 1240, 2480 and 4960 ppm $CO_2$. However, about 25 percent of actual paleoclimate GHG forcing is likely supplied by non-$CO_2$ GHGs, in which case these experiments should be interpreted as having 525, 900, 1570 and 2755 ppm of $CO_2$.

    Simulations with the Russell climate model yield poleward amplification of the warming as the forcing increases (Fig. S6), but, as in other climate models, the high latitude warming is much less than in paleoclimate data. This model deficiency has been discussed extensively with regard to the Pliocene (Dowsett et al., 2009; Lunt et al., 2012a) and Eocene (Greenwood and Wing, 1995; Huber and Caballero, 2011; Lunt et al., 2012b). Although various suggestions have been made for physical mechanisms that may be missing or poorly represented in the models, we suspect that much of the problem is a consequence of unrealistic poleward transport of heat by the ocean in most models. We presented evidence recently (Hansen et al., 2011) that most ocean models are too diffusive, which could affect dynamical transports. Recent Pliocene simulations with a version of the Goddard Institute for Space Studies model that uses improved parameterizations of sub-grid-scale mixing, for example, have a northward shift and 25 percent increase in the strength of the Atlantic overturning circulation (Chandler et al., 2012). Simulations in our present paper are made with an ocean model having a depth of only 100 m, and thus we have no expectation that ocean poleward heat transports can be realistic. We plan to repeat some simulations with a full ocean model, but the computation time will be greatly increased and such further simulations are beyond the scope of our present paper.

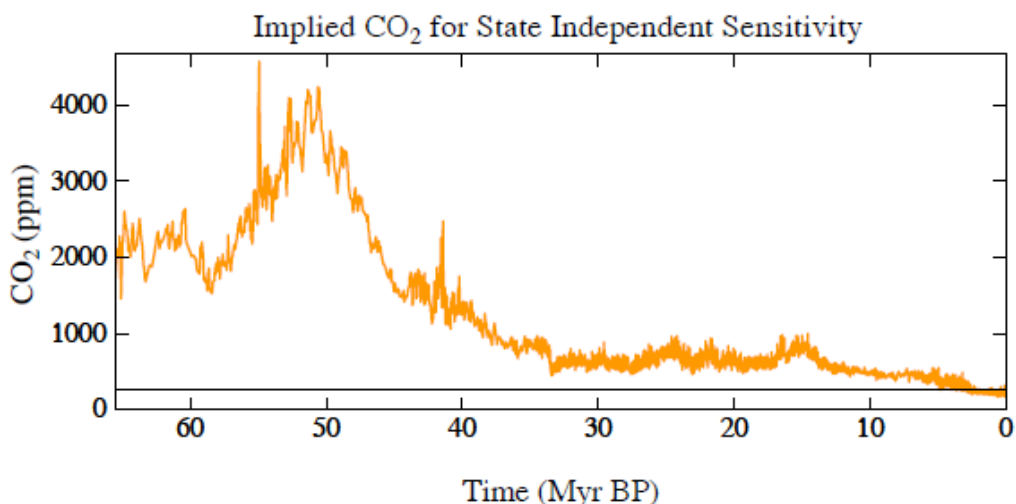

**Fig. S7.** Same as Fig. 8a, except we have here assumed that global temperature during the Early Eocene Climatic Optimum was ~33°C rather than the ~28°C that was assumed in calculations for Fig. 9a.

In the calculations for Figs. S7, S8 and S9 the pre-PETM global temperature is ~30.2°C, 4.5°C warmer than for the calculations that produced Figs. 8, 9 and 10. The principal implication of Figs. S7, S8 and S9 is that a PETM warming of ~5°C favors a fast-feedback climate sensitivity close to the full Russell sensitivity.

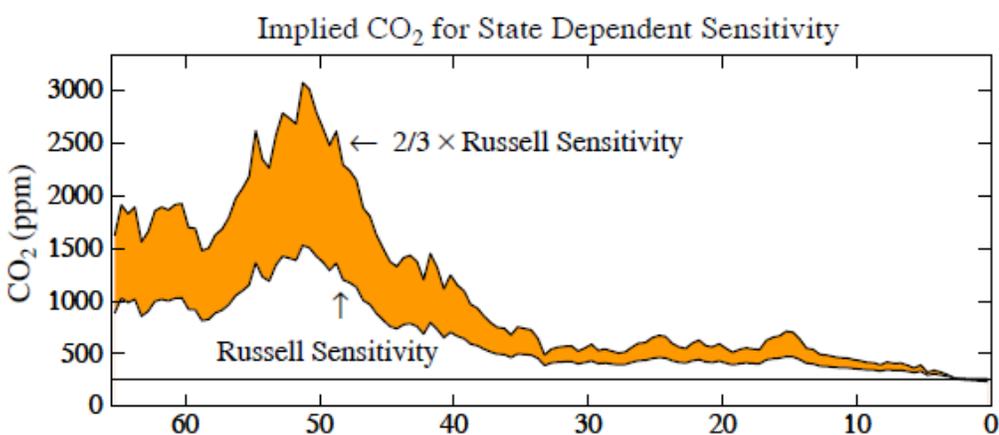

**Fig. S8.** Same as Fig. 9a, except we have here assumed that global temperature during the Early Eocene Climatic Optimum was ~33°C rather than the ~28°C that was assumed in calculations for Fig. 9a.

Fig. S9 shows that 1200 ppm $CO_2$ is required to yield a PETM warming of 5°C for the Russell climate sensitivity, but 3500 ppm would be needed if sensitivity is only 2/3 of the Russell sensitivity. Given that the estimated PETM carbon injection of 4000-7000 PgC (see main text) corresponds to ~630-1100 ppm $CO_2$, the higher Russell sensitivity is required to achieve reasonably good consistency with PETM warming, if the maximum Eocene temperature was ~33°C.

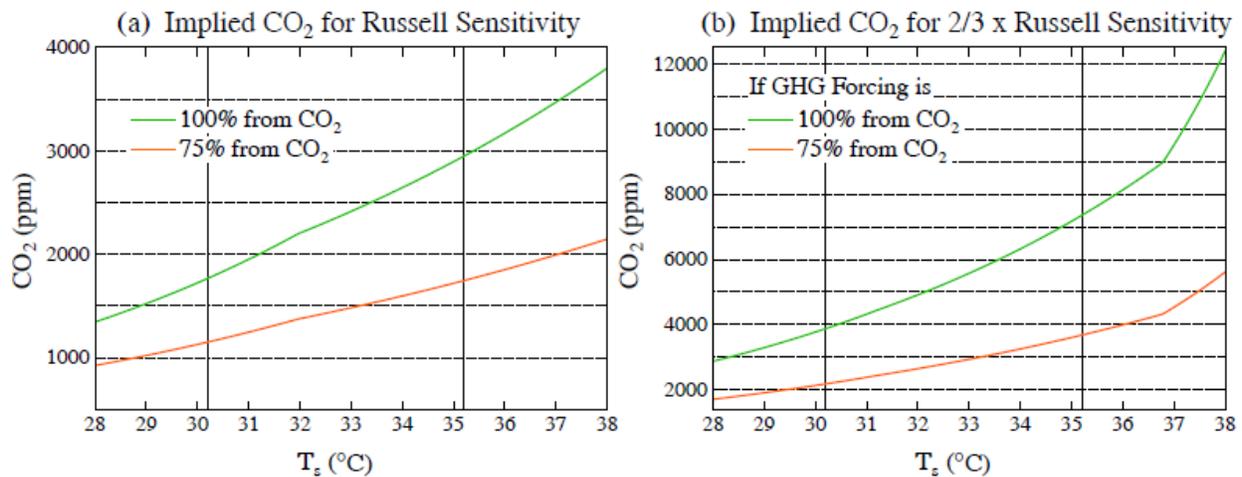

**Fig. S9.** Same as Fig. 10, except we have here assumed that global temperature during the Early Eocene Climatic Optimum was ~33°C, and thus the pre-PETM temperature was ~30.2°C.